\documentclass[manuscript,screen]{acmart}

\usepackage{kantlipsum}
\usepackage{colortbl}
\usepackage{multicol}
\usepackage{graphicx}
\usepackage{caption,subcaption, ragged2e}
\usepackage{amsmath}
\usepackage{comment}
\usepackage{xspace}

\usepackage[british]{babel}
\usepackage{hhline}
\usepackage{multirow}

\usepackage{booktabs}
\usepackage{multirow}
\usepackage{siunitx}

\usepackage{array}
\usepackage{makecell}
\usepackage{graphicx}
\usepackage{multirow}
\usepackage{colortbl}
\usepackage{color}
\usepackage{tabularray}
\usepackage{tabularx}

\usepackage{pifont}
\newcommand{\xmark}{\textcolor{red}{\ding{55}}}%

\usepackage{longtable}

\usepackage{xcolor,pifont}
\newcommand*\colourcheck[1]{%
  \expandafter\newcommand\csname #1check\endcsname{\textcolor{#1}{\ding{52}}}%
}
\colourcheck{blue}
\colourcheck{green}
\colourcheck{red}

\AtBeginDocument{%
  }

\usepackage[marginparwidth=2cm]{}

\setcopyright{rightsretained}
\copyrightyear{2024}
\acmYear{2024}
\acmDOI{XXXXXXX.XXXXXXX}

\acmBooktitle{ACM Transactions on Cyber-Physical Systems, Vol. xx, No. x, Article xx. Publication date: xxxx xxxx.}
\acmISBN{978-1-4503-XXXX-X/18/06}

\begin{document}
\newcommand{\Name}{MARS}
\title{MARS: \underline{M}ulti-radio \underline{A}rchitecture with \underline{R}adio \underline{S}election using Decision Trees \\
for emerging mesoscale CPS/IoT applications}

\author{Jothi Prasanna Shanmuga Sundaram}
\email{jshanmugasundaram@ucmerced.edu}
\affiliation{%
  \institution{University of California, Merced}
  \state{California}
  \country{USA}
}

\author{Arman Zharmagambetov}
\email{armanz@meta.com}
\affiliation{%
  \institution{FAIR, Meta AI}
  \city{Menlo Park}
  \state{California}
  \country{USA}
}

\author{Magzhan Gabidolla}
\email{mgabidolla@ucmerced.edu}
\affiliation{%
  \institution{University of California, Merced}
  \state{California}
  \country{USA}
}

\author{Miguel \'A.\ Carreira-Perpi\~n\'an}
\email{mcarreira-perpinan@ucmerced.edu}
\affiliation{%
  \institution{University of California, Merced}
  \state{California}
  \country{USA}
}

\author{Alberto Cerpa}
\email{acerpa@ucmerced.edu}
\affiliation{%
  \institution{University of California, Merced}
  \state{California}
  \country{USA}
}

\renewcommand{\shortauthors}{}

\begin{abstract}
Internet-of-Things is rapidly growing from traditional small-scale applications like smart homes (10-100m range) to large-scale applications like Microsoft Farmbeats (1-5Km range). Small-scale applications are catered by short-range radios like Zigbee and Bluetooth LE while large-scale applications are catered by long-range radios like LoRa and NB-IoT. The other upcoming category of IoT applications like P2P energy-trade in smart homes (0.1Km-1Km range) are termed mesoscale IoT applications. There are no specialized radios fully developed for mesoscale IoT applications. They either use short-range or long-range radios. To close this gap, we explored mesoscale IoT applications using commercial-off-the-shelf (COTS) IoT radios available. Our qualitative analysis identifies Zigbee and LoRa as potential candidates. Further quantitative analysis of radio candidates on both single-hop and multi-hop topologies showed that Zigbee and LoRa achieve competitive throughput at the distance of 500-1200m from the gateway, termed the \textit{gray-region}. A fundamental finding of these analyses is that a multi-radio system that can efficiently switch between Zigbee and LoRa performs better than the single-radio systems. However, instantaneously selecting and switching to a high-throughput radio at the time of transmission is not trivial because of erratic link quality dynamics in mesoscale IoT environments.

To address this issue, we developed \Name: a Multi-radio Architecture with Radio Selection using Decision Trees that use instantaneous end-to-end path quality metrics to instantaneously select the high-throughput radio at the time of transmission. However, realizing \Name\xspace on resource-constrained end-devices entails the challenge of obtaining instantaneous path quality information. Traditional path quality estimation is not instantaneous due to propagation and queuing delays. We overcome this challenge by showing that collecting local path metrics as input to our decision trees provides sufficient information to instantaneously identify the high-throughput radio. The radio selector of \Name\xspace is powered by TAO-CART trees. These trees are converted into IF...ELSE statements for efficient deployment on end devices. The evaluation of \Name\xspace on a large-scale mesh topology at two different locations show that \Name\xspace can efficiently identify and switch to the high-throughput radio at the time of transmission, leading to an average throughput gain of 48.2\% and 49.79\% than the competing schemes\footnote{This article has been submitted to ACM Transactions on Cyber-Physical Systems for consideration. Its contents may not be further disseminated until a final decision regarding publication has been made and further permission has been granted}.
\end{abstract}




\maketitle

\section{Introduction}
Traditional IoT networking comprises both small-scale applications, like smart homes (10-100m range), and large-scale applications, like Microsoft FarmBeats~\cite{vasisht2017farmbeats} (1-5Km range). On the other hand, emerging applications serve in the range of 0.1-1.5Km, like smart-grid Neighborhood Area Networking (NAN)~\cite{ding2019constrained}, target tracking~\cite{yeow2007energy}, Industrial Automation \cite{ojo2018throughput, zhang2019throughput} and 
P2P energy-trade in smart neighborhood~\cite{pilz2017recent, saif2022impact, liu2019intraday}. These applications are hereafter referred to as \textit{mesoscale IoT applications}.
In P2P energy trade \cite{mitsubishielectricSmartMeter}, multiple nodes bid for available energy like stock market transactions. Industrial automation \cite{ mitsubishielectricLargecapacityBattery, mitsubishielectricPowerSystems, ma2022smart, fu2024comprehensive} provides time-critical closed-loop control for M2M communications. Here, the radios should optimize latency/throughput for better application performance \cite{sisinni2018industrial}. The latency/throughput are calculated on successfully delivered packets, indirectly including reliability. 
The IoT radio utilized here will be a small component of a larger device. Unlike the typical IoT nodes powered by a smaller battery, these devices are either grid-powered \cite{yang2024rateless, nagai2024improve} or have a large battery reserve to support the entire device \cite{mitsubishielectricSmartMeter, mitsubishielectricLargecapacityBattery, mitsubishielectricPowerSystems}. 
Hence, we chose latency/throughput as the performance indicator with a secondary emphasis on energy consumption.  

Unlike the comparatively stable deployed environments of smart homes and Microsoft Farmbeats~\cite{vasisht2017farmbeats}, mesoscale applications serve in urban/semi-urban environments that are relatively dynamic. This dynamic nature is because of environmental obstacles, like buildings made of wood, glass, and concrete, heavy human influx, and vehicle movements. These factors erratically alter the wireless link quality. 
Some applications may need direct single-hop communication with the gateway while other applications may need multi-hop communication to communicate with other nodes in the network. 
Since LoRa covers the entire mesoscale range, an intuitive idea would be to use LoRa in a multi-hop fashion, for applications that need inter-node communications.
However, using a low-throughput LoRa radio in a multi-hop fashion will further reduce the end-to-end throughput with the increase in hops~\cite{boorstyn1987throughput}.
Furthermore, mesoscale IoT applications do not have a dedicated, well-developed radio technology. 
To close this gap, we explored the performance of the available Commercially-Off-The-Shelf (COTS) IoT radios for the mesoscale IoT applications.

First, we characterized the mesoscale IoT environment and conducted a qualitative analysis on the available IoT radios~(\S\ref{sec:lpradios}). This analysis identifies Zigbee and LoRa as potential candidates for mesoscale IoT applications. Further quantitative analysis (refer \S\ref{subsec:perf_analysis}) of the radio candidates on both single-hop and multi-hop topologies show that Zigbee and LoRa achieve competitive throughput at the distance of 500-1200m from the gateway, hereafter referred as the \textit{gray-region}. 
In this \textit{gray-region},
LoRa and Zigbee radios achieve high end-to-end throughput at different time instants due to erratic channel conditions ( refer \S\ref{subsec:e2e_fluc}).
Unlike transitional region\cite{zuniga2004analyzing} that shows the dynamicity of PDR between connected and disconnected regions of a single hop, the gray region shows that two radios with an order of magnitude difference in theoretical throughput, achieve competitive throughput, over multiple hops.
The fundamental finding is that Zigbee and LoRa can work together as a multi-radio system and efficiently switch between the radios to maximize throughput (refer \S\ref{sec:motivation}).
However, instantaneously selecting a high-throughput radio is not a trivial problem because of erratic link quality dynamics in the mesoscale IoT application environments. 

Second, we developed \Name\xspace to instantaneously select high throughput radio during transmission, using instantaneous end-to-end path/link quality metrics. \Name's node comprises both Zigbee 2.4 GHz and LoRa 915 MHz radios. While it is intuitive to employ a Machine Learning (ML) model to perform radio selection, it entails the challenge of obtaining instantaneous path/link quality estimations.
Multi-hop Zigbee radios propagate the link quality information of all the links along the path to compute path quality~\cite{de2003high}. However, the propagated link quality info expires before it reaches the destined node because of temporal link quality variations, propagation, and queuing delays along the path~(refer \S\ref{sec:trad_not_inst}). While it is ideal to compute the path quality of the entire path, we observed that a part of the entire path length is sufficient to identify the high throughput radio during transmission. We performed extensive analysis and evaluation on multiple topologies, to find the fewer hops required to estimate the end-to-end path quality as input to our ML model and still obtain good radio selection accuracy (refer \S\ref{subsec:DT_est}). This way we balance the trade-off between perfectly accurate global metrics that cannot be gathered on time and acceptably accurate local metrics that can be collected on time while still providing good input for our ML model, trained by CART\cite{praagman1985classification}, to achieve acceptable 
radio selection 
accuracy. 

Finally, we optimize the CART with Tree-Alternating Optimization (TAO)~\cite{carreira2018alternating}. The TAO algorithm optimizes the traditional trees to achieve higher accuracy with lower training data requirements. The TAO-Optimized CART (TAO-CART) can be converted into IF...ELSE statements for efficient deployment on IoT end devices. An evaluation of \Name\xspace on a complex mesh topology at two different locations showed that \Name\xspace can instantaneously select the high-throughput radio during transmission leading to an average throughput gain of 48.2\% and 49.79\% than the competing schemes. 

In summary, the contributions of our work are:
\begin{enumerate}
    \item Identifying the absence of a dedicated radio for emerging mesoscale IoT apps, we developed an intelligent multi-radio architecture with COTS IoT radios, namely Zigbee and LoRa, that has an order of magnitude difference in theoretical throughput.
    \item We identified the existence of a \textit{gray-region} between 0.5-1.2 Km from the gateway, through analytic and experimental analysis, showing that due to the temporal variability in the channel, it is uncertain which radio provides the best throughput at any given time.
    \item We developed an implementation of an ML Decision Tree model for radio selection using the TAO~\cite{carreira2018alternating} algorithm.  To the best of our knowledge, this is the first real-world use case of TAO.
    \item We showed that partial path quality can provide sufficient information for our TAO-optimized tree to accurately select the high-throughput radio during transmission.
\end{enumerate}

\section{Related work}
\label{sec:Related-Work}

Multi-radio wireless networks have been heavily worked on for a decade using different radio combinations. WiFi + LTE~\cite{bennis2013cellular, deng2014wifi}, WiFi + Bluetooth~\cite{agarwal2008switchr, ananthanarayanan2009blue, pering2006coolspots}, 60GHz+WiFi~\cite{sur2017wifi} to ameliorate energy efficiency~\cite{agarwal2008switchr, jin2011wizi, kusy2014radio, lymberopoulos2008towards}, traffic management~\cite{ferlin2014multi}, mobility management~\cite{pluntke2011saving} and routing management~\cite{alicherry2005joint, draves2004routing}. Bahl et al.~\cite{bahl2004reconsidering} identified that multi-radio systems are beneficial for wireless networks. They showed that using different radio technologies on the same platform improves performance. They recommended abstracting multiple radios as a single logical pipeline yet providing access to the networking protocols. While most of the above multi-radio systems were developed for high-power radios like WiFi, the systems designed for IoT networks are discussed below.

Backpacking~\cite{al2011backpacking} was developed for high data rate sensor networks like HP's CeNSE~\cite{hpcense}. Initially, they experiment with the network-level energy efficiency of the low-power (802.15.4) and high-power (802.11b) radios on different data rates, node densities, and distances. Using these results, a cross-layer empirical model is developed to calculate the optimal density of high-power radios to be augmented with the low-power radios, to strike a balance between the usage of the two radio types. They use 802.15.4 and 802.11b radios in the accumulator node and only 802.15.4 radio in the originator node. Originator nodes send smaller packets of sensed data to the accumulator node using 802.15.4 radio. 
The accumulator acts as a relay between multiple originator nodes and the base station. An accumulator node gathers all the data to be sent to the base station from multiple originator nodes and uses 802.11b radio to transmit the accumulated data to the base station. Their indoor experiments inside a small room show that Backpacking can gain 44\% improvement in average energy per bit. This will work perfectly for high-data-rate, non-deadline-oriented applications but won't work for low-data-rate, time-sensitive applications because an accumulator node has to gather a lot of data before transmitting via 802.11 radio. Unlike this work, we develop a multi-radio architecture for low data rate, time-sensitive mesoscale IoT applications. 

Kusy et al.~\cite{kusy2014radio} develop a multi-radio architecture for wireless sensor networks. They show that employing two multi-hop radios in the same node improves communication reliability but pays a toll of 3-33\% higher energy consumption. They exploit radio diversity with two 802.15.4 radios working in mutually exclusive bands of 2.4 GHz and 900 MHz to enable simultaneous transmission and reception capabilities. They abstract the radios behind a software driver that enables simultaneous, independent control and data plane operations. Their experiment on a 400m$\times$450m area achieves communication reliability but pays a 33\% overhead in energy consumption as multi-hop networks will incur more delay and energy when more hops are added~\cite{al2011backpacking}. Unlike this work, we focus on optimizing throughput and latency for mesoscale IoT applications, ranging from 100-1000m from the gateway. Hence, it is clear that using two independent, simultaneously operating multi-hop networks for a mesoscale IoT application is not energy-inefficient and they incur more delay at each hop leading to a reduced throughput. These results suggest to use one multi-hop radio and one single-hop radio with diverse frequency ranges to improve the performance of mesoscale IoT applications.  

Gummeson et al.~\cite{gummeson2009adaptive} found that range diversity and channel dynamics are the major impediments to energy efficiency in mobile sensor networks.  
They developed a Reinforcement Learning (RL) based adaptive link layer to predict channel dynamics. The reward of this RL algorithm is the expected energy efficiency when choosing a particular radio for transmission. On learning the channel dynamics, they employ a formula to translate the learned channel dynamics to estimate the energy required by each radio for transmitting a packet. Finally, they choose a radio that consumes less energy. They also propose a protocol to coordinate the data plane and control plane operations of the two radios. The RL algorithm is the appropriate choice for learning channel dynamics of complex mobile sensor networks, but their complexity is very high (refer \S\ref{sec:eval} (ii)) for static IoT network employed for mesoscale IoT applications.  

Lymberopoulos et al.~\cite{lymberopoulos2008towards} provided guidelines to design a multi-radio architecture for energy-efficient Wireless Sensor Networks (WSN). They use 802.11b and 802.15.4 radios in the same platform and study the effect of packet size, and transmission time on energy consumption. Through experiments, they identified the bottlenecks, trade-offs, and break-even points of multi-radio systems. Based on these break-even points,
they developed a threshold-based radio-switching algorithm to optimize energy efficiency. This threshold-based algorithm utilizes the break-even point regions, where one radio performs better than the other, to switch to an energy-efficient radio.
\begin{table*}[t]
\centering
\caption{A comparison of related work}
\label{tab:related-comp}
\resizebox{\textwidth}{!}{
\begin{tabular}{|c|c|cccc|c|c|c|}
\hline
\multirow{2}{*}{Multi-radio systems} &
  \multirow{2}{*}{Radios Used} &
  \multicolumn{4}{c|}{Optimized Metric} &
  \multirow{2}{*}{Deployment Range  \textgreater{}500m?} &
  \multirow{2}{*}{Radio Selector} &
  \multirow{2}{*}{Mobility?} \\ \cline{3-6}
 &
   &
  \multicolumn{1}{c|}{Energy} &
  \multicolumn{1}{c|}{Reliability} &
  \multicolumn{1}{c|}{Throughput} &
  Latency &
   &
   &
   \\ \hline
Backpacking~\cite{al2011backpacking}&
  802.11 + 802.15.4 &
  \multicolumn{1}{c|}{\greencheck} &
  \multicolumn{1}{c|}{\xmark} &
  \multicolumn{1}{c|}{\xmark} &
  \xmark &
  \xmark &
  \xmark &
  \xmark \\ \hline
Kusy   et al.~\cite{kusy2014radio}&
  802.15.4   (915MHz) + 802.15.4 (2.4GHz) &
  \multicolumn{1}{c|}{\xmark} &
  \multicolumn{1}{c|}{\greencheck} &
  \multicolumn{1}{c|}{\xmark} &
  \xmark &
  \xmark &
  \xmark &
  \xmark \\ \hline
Gummeson   et al.~\cite{gummeson2009adaptive}&
  802.15.4   (915MHz) + 802.15.4 (2.4GHz) &
  \multicolumn{1}{c|}{\greencheck} &
  \multicolumn{1}{c|}{\xmark} &
  \multicolumn{1}{c|}{\xmark} &
  \xmark &
  \xmark &
  Reinforcement   Learning &
  \greencheck \\ \hline
Lymberopoulos   et al.~\cite{lymberopoulos2008towards} &
  802.11b   + 802.15.4 &
  \multicolumn{1}{c|}{\greencheck} &
  \multicolumn{1}{c|}{\xmark} &
  \multicolumn{1}{c|}{\xmark} &
  \xmark &
  \xmark &
  Threshold-based &
  \xmark \\ \hline
MARS &
  802.15.4 (2.4GHz) + LoRa LPWAN &
  \multicolumn{1}{c|}{\xmark} &
  \multicolumn{1}{c|}{\xmark} &
  \multicolumn{1}{c|}{\greencheck} &
  \greencheck &
  \greencheck &
  Decision   Trees &
  \xmark \\ \hline
\end{tabular}%
}
\vspace{-0.3cm}
\end{table*}

A comparison of related work with MARS is tabulated in Table~\ref{tab:related-comp}. While all the previous works employed multi-radio systems for energy efficiency and improving reliability, MARS is optimizing throughput and latency. To the best of our knowledge, MARS is the first to explore multi-radio architecture for mesoscale applications ranging between 100-1000m from the gateway. 
Two closely related works are Lymberopoulos et al.~\cite{lymberopoulos2008towards} and Gummeson et al.~\cite{gummeson2009adaptive}. We made our best effort to adopt the threshold-based algorithm of Lymberopoulos et al.~\cite{lymberopoulos2008towards} for comparison with MARS.  The sophisticated RL-based radio switching protocol of Gummeson et al.~\cite{gummeson2009adaptive} suffers from the below-described problems: (i) During data transfer between two radios in the communication range, the radio switching protocol, designed for energy efficiency, does a three-way handshake to switch radios. This incurs additional latency which will heavily degrade the throughput. (ii) A well-known issue of model-free RL is that it requires heavy training data to converge to an acceptable performance~\cite{ding2020mb2c} and the amount of data used for training their model-free RL model is obscure. We compared the training data requirements of Q-learning and Decision Tree models in \S\ref{subsec:ML_models}. Since Gummeson et al.~\cite{gummeson2009adaptive} developed custom hardware for energy efficiency, we made our best effort to adapt their work to compare with MARS. 

\begin{figure*}[t]
  \vspace{-0.1cm}
  \resizebox{\textwidth}{!}{
  \includegraphics[width=0.7\textwidth, height=3.7cm]{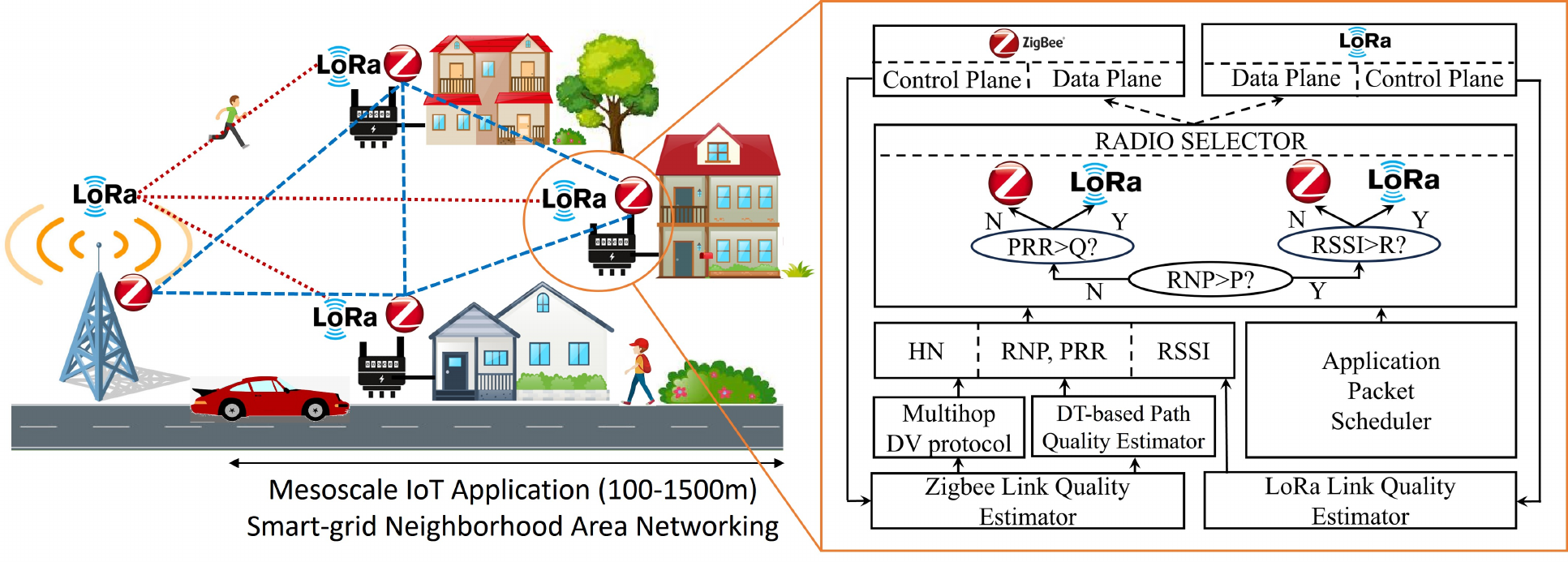}}
  \vspace{-0.5cm}
  \caption{MARS System Overview}
  \label{fig:overview}
  \vspace{-0.3cm}
\end{figure*}
\section{\Name\xspace System Overview}
\label{sec:overview}

In this section, we provide an overview of \Name\xspace. Figure~\ref{fig:overview} shows the system overview with all the components. \Name\xspace consists of a network of nodes using multiple radios, chosen using the methodology explained in \S~\ref{sec:lpradios}. Multi-hop Zigbee network uses a
distance-vector routing~\cite{cheng1989loop}
protocol for multi-hop routing. The control plane of both radios constantly feeds the link quality estimators. LoRa link quality estimator directly feeds the Radio selector with RSSI. The DT-based path-quality estimator takes input from the Zigbee LQ estimator to estimate path qualities RNP and PRR.  The specific path quality metrics used are not completely end-to-end. The analysis and determination of those metrics are explained in \S~\ref{subsec:DT_est}, where we show that a partial path quality is sufficient to provide inputs to the ML model to achieve acceptable radio-selection accuracy, while at the same time being able to gather this data in a fast enough manner as to preserve the temporal correlations in the RF channel. The ML model uses an input feature vector composed of multiple path and link quality metrics explained in \S~\ref{subsec:Problem_formulation}. The reason for using an ML model is provided in \S~\ref{subsec:e2e_fluc}, and it is related to the temporal dynamics in the gray-region explained in \S\ref{subsec:e2e_fluc}.  Taking these inputs, the radio selector powered by an ML model, explained in \S~\ref{sec:ML_build}, chooses the high throughput radio at the time of transmission. 

\section{Factors affecting mesoscale IoT applications}
\label{subsec:factors}

A mesoscale IoT environment ranges anywhere between 100-1500m~\cite{ding2019constrained, yeow2007energy, sastry2006instrumenting}. The factors affecting the performance of mesoscale IoT applications are three-fold: (i) Environmental obstacles, (ii) Vagarious IoT radios, and (iii) Applications demanding different communication paradigms and Quality of Service (QoS). 

Firstly, a mesoscale IoT environment will be very dynamic in nature. 
Mesoscale IoT environment comprises buildings, heavy human influx, and vehicle movements. The buildings are made of materials like steel, glass, concrete, physical firewalls, etc. Wireless signals passing through these objects get attenuated to a fixed level that can be approximated with a mathematical model. Whereas, humans and moving objects, like vehicles, attenuate the wireless signals in an erratic manner. Capturing these erratic variations through a mathematical model is complicated and overwhelming. 

Secondly, radios developed for IoT applications are designed to be very cheap with low energy consumption so that they can be deployed in thousands of IoT devices. These low-cost IoT radios are vagarious. For example, a LoRa radio slightly deviates from its central frequency during transmission~\cite{sundaram2019survey}. These vagarious IoT radios, deployed in a mesoscale IoT environment that causes various levels of signal attenuation, will fluctuate the performance of IoT networks in an unpredictable manner.    


Thirdly, with the rise of multi-tenancy \cite{cherrier2015multi} in IoT networks, where each end-node serve multiple applications, each application may need different communication paradigms. For example, In an asset monitoring application, a node reports the status of an asset to the base station periodically. A single-hop link to the base station is more suitable for this application than the multi-hop paradigm. Applications like target-tracking need multi-hop inter-node communications to coordinate with other devices for efficiently tracking the target and actuating necessary functions. LoRa works better for applications like asset monitoring and Zigbee works better for applications like target tracking.

Furthermore, for applications requiring high data rate bi-directional communications, WiFi-HaLow will achieve higher performance than LoRa as LoRa is precisely designed for up-link oriented applications~\cite{rizzi2017using}. Since each radio has its own merits and demerits, using single-radio networks to cater to all the applications in the above-described dynamic mesoscale IoT environment is debatable. Hence, it is crucial to analyze the characteristics and performance of different IoT radios for mesoscale IoT applications. 

\section{LP-Radios for mesoscale applications}
\label{sec:lpradios}

In pursuit of finding a radio for mesoscale IoT applications, we first qualitatively compare the suitability of different IoT radios for mesoscale IoT applications. On identifying suitable radio candidates, we further conduct quantitative performance analysis to understand their strengths and pitfalls.

\label{subsec:radio_comparison}
\textbf{A qualitative comparison.} 
\begin{table}[t]
  \centering
  \vspace{-0.1cm}
  \caption{A comparison of different IoT radios in the context of city-scale smart environments showing that Zigbee and LoRa are the better-suited radios.}
  \label{tab:iot_radio_comparison}
  \vspace{-0.2cm}
  \begin{tabular}{cccccc}
    \toprule
                     & SigFox           & WiFi HaLow                        & BLE                & Zigbee                 & LoRa\\  
    \toprule
    Open-source?        & \cellcolor{red}No&\cellcolor{green}Yes&\cellcolor{green}Yes&\cellcolor{green}Yes    &\cellcolor{green}Yes\\
    Link Budget (dB)    & 158~\cite{SigFox}&\cellcolor{red}24.5~\cite{lee2021wifi}&\cellcolor{green}108~\cite{le2017bluetooth}&\cellcolor{green}103~\cite{Zigbee}&\cellcolor{green}150~\cite{petajajarvi2017evaluation}\\
    Topology Type                & LPWAN            &LPWAN       &\cellcolor{red}PAN&\cellcolor{green}PAN/LAN&\cellcolor{green}LPWAN\\
    Communication Range (m)           & 5000             &1000        &100     &\cellcolor{green}125~\cite{milenkovic2005environment}&\cellcolor{green}5000\\
    Max bitrate (bps)   & 600              &upto 4M     & up to 1M      &\cellcolor{green}250K   &\cellcolor{green}upto 27K\\ 
   \bottomrule
   \vspace{-0.2cm}
  \end{tabular}
  \vspace{-0.2cm}
\end{table}
A comparison of different IoT radios is tabulated in Table~\ref{tab:iot_radio_comparison}. SigFox is not an open-source technology. Hence, it cannot be used for deploying private networks. WiFi HaLow's (802.11ah) low link budget won't allow the signals to penetrate through the environmental obstacles. Hence, SigFox and WiFi-HaLow do not suit mesoscale IoT applications. Zigbee is a multi-hop, high-bit rate radio with a moderate link budget to penetrate environmental obstacles. Single-Hop LoRa is also a better fit as it has a strong link budget satisfying all the requirements of mesoscale IoT applications. Zigbee and LoRa are chosen for further quantitative analysis.

\subsection{A quantitative comparison of IoT radios}
\label{subsec:perf_analysis}

Zigbee and LoRa are identified as potential radio candidates for further quantitative analysis to choose a suitable radio for mesoscale IoT applications. In this analysis, the performance of both radio candidates is compared by their throughput and Packet Loss Ratio (PLR) in free space and an urban-like mesoscale environment. This analysis is done for both single and multi-hop topologies for a Zigbee radio.   

\subsubsection{\textbf{Single-hop experiments.}} A sender and receiver of both the radio candidates are placed in both free space and an urban-like mesoscale environment at a distance of 20m from each other. The latter spans multiple wooden walls, a glass door, and three humans in between the nodes. One thousand packets are transmitted from the sender to the receiver to average the achieved throughput and PLR. 

\textbf{Throughput of Zigbee 2.4GHz.}
One thousand 29-byte-sized packets are sent back-to-back without any delay. Another receiving mote logs the received packets. CSMA and link-layer acknowledgments are disabled in TinyOS. There is no reliable mechanism for the retransmission of packets. Zigbee achieves a 100\% packet reception ratio with an average throughput of 77,634 bps in free space. The theoretical maximum of 250 Kbps is not achieved as throughput heavily depends on packet size~\cite{ganti2006datalink}. In an urban-like environment, 56,530 bps throughput was achieved with 33.40~\% PLR. Our experiments closely matched the results by Opal~\cite{jurdak2011opal}, Hamdy et al.~\cite{r2020evaluation} and Yousuf et al.~\cite{yousuf2018throughput}. The difference between free-space and urban-like environments is the increase in packet loss ratio and a reduction in throughput. Heavy packet loss is due to the fact that the signals are either lost or corrupted because of the urban-like environment. The reduction in throughput is because the signals are penetrating through the obstacles in an urban-like environment. Any retransmission mechanism incorporated will reduce the overall network throughput. This packet loss is due to the fact that signals transmitted in 2.4GHz have moderate penetration capacity and were not received/decoded by the receiver. An intuitive idea would be to utilize Zigbee radio at a lower frequency of 915 MHz for an increased signal penetration capacity.  

\begin{table*}[t]
  \vspace{-0.1cm}
  \caption{Throughput and packet loss rate of Zigbee 2.4GHz, Zigbee 915 MHz and LoRa 915 MHz showing that Zigbee 2.4GHz is better for short-range communications and LoRa 915MHz is better for long-range communications in smart-neighborhood environment.}
  \label{tab:radio_comparison}
  \vspace{-0.2cm}
  \centering
  \begin{tabular}{ccl}
    \toprule
    Radio candidate & Throughput (bps) & PLR (\%)\\  
    \toprule
    Zigbee 2.4 GHz in free space & 77,634 & 0 \\ 
    Zigbee 2.4 GHz in smart environment & 56,530 & 33.40 \\
    Zigbee 915 MHz in free space  & 9,530 & 0 \\
    Zigbee 915 MHz in smart environment & 6,777 & 24.60 \\
    LoRa 915 MHz in free space & 4,579 & 0\\
    LoRa 915 MHz in smart environment & 4,579 & 0 \\ [1ex] 
    \bottomrule
    \vspace{-0.5cm}
  \end{tabular}
  \vspace{-0.3cm}
\end{table*}
\textbf{Throughput of Zigbee 915MHz and LoRa 915MHz}. Zigbee915 achieves a packet reception ratio of 100\% in free space with an average throughput of 9530 bps. An average throughput of 6777 bps is achieved with 24.60\% PLR in a mesoscale environment. As the loss is higher in Zigbee 915 MHz, we wanted to compare a relatively newer LoRa radio working in 915 MHz with Zigbee 915 MHz. Two LoStik~\cite{ronoth_llc} LoRa USB nodes using SF7 in 125KHz bandwidth are placed in free-space and urban-like environments at a distance of 20m in between them. One thousand 29-byte-sized packets were sent from one node to another. LoRa was able to achieve 4579 bps average throughput with a 100\% reception ratio in both free-space and smart-neighborhood environments. The penetration capacity of Zigbee 915MHz cannot match LoRa's penetration capacity as LoRa generates higher link-budget signals. 
From the above single-hop experiments, it is evident that Zigbee 2.4 GHz achieves higher throughput than Zigbee 915MHz. 
This high throughput characteristic is highly desirable for mesoscale applications despite the higher loss rate since the throughput will degrade further in multi-hop communications owing to queuing and channel sensing delays~\cite{boorstyn1987throughput}. If we choose Zigbee
915 MHz because of a lower loss rate, its lower throughput will further degrade when employed in a multi-hop fashion. So, we choose Zigbee 2.4GHz over Zigbee 915MHz. Comparing LoRa 915MHz and Zigbee 915MHz, LoRa 915 MHz achieves lower PLR than Zigbee 915 MHz as LoRa signals are robust owing to its CSS modulation scheme~\cite{sundaram2019survey}. Hence, it is identified that Zigbee 2.4GHz is the better radio for mesoscale IoT apps demanding short-range, multi-hop communications while LoRa is better for mesoscale IoT applications demanding long range, single-hop communications.

\subsubsection{\textbf{Multi-hop experiments}}
\label{subsubsec:multiho_exp}
The next step is to test the performance of Zigbee and LoRa in a simple multi-hop line topology. Multi-hop line topology might affect the throughput of Zigbee due to channel sensing (CSMA) and queuing (forwarding) delays.    
The multi-hop experiments are conducted in two folds. First, an analytic performance analysis is conducted using the available models. Second, the result of this analytic performance analysis is compared with the real-world experimental performance analysis. 

\textit{\textbf{Topology setup.}}
\label{topo_setup}
    A simple line topology is used for this analysis. One gateway and fifteen nodes are placed in line topology in free space such that each hop spans approximately 100m. Packets of size 29 bytes are generated according to an Independent Poisson Process at the rate of 1 packet every 3 seconds. Each node has both LoRa and Zigbee radios. All generated packets are destined for the gateway. LoRa can reach the gateway in a single hop whereas Zigbee takes multiple hops. LoRa nodes transmit in a 125KHz channel with an SF7 spreading factor. LoRa gateway is capable of receiving eight packets concurrently~\cite{sundaram2019survey}. This highly reduces collision.  
    
\textbf{Analytic throughput analysis.}
\label{analytics}
LoRa uses ALOHA where nodes can transmit at their will. Packets are transmitted at the Poisson rate $\lambda$ packets/second assuming that each packet occupies the channel for $\tau$ seconds. Then the normalized traffic G = $\lambda\times\tau$. A packet will be corrupted due to channel noise or when two packets are transmitted at the same time. The former case is ignored in this analysis. The occurrence of the correctly received packets can be defined as $\lambda' < \lambda$. Then the normalized channel throughput is given by S = $\lambda'\times\tau$. Packet reception is successful only if there is no other transmission in the interval [$-\tau, \tau$]. Since the Poisson process defines all the transmission times, the probability that two packets won't collide is $e^{-2\lambda.\tau}$ = $e^{-2G}$. The throughput is given by multiplying the normalized channel throughput and the probability of successful packets~\cite{abramson1977throughput}. 
Therefore the throughput of LoRa is given by $S = Ge^{-2G}$.

Zigbee network employs CSMA. Each node schedules packets to its neighbors according to an independent Poisson point process. By assuming an independent Poisson process, relaying and queuing delays can be ignored. This network can be modeled using a Continuous Time Markov Chain with a set of nodes transmitting at any time instant~\cite{boorstyn1987throughput}. Let i be a node. $N_{i}$ be the set of all neighbors of i. Let $g_{ij}$ and $s_{ij}$ be the scheduled and desired packet rates when a packet is transmitted from node i to j. Let P(A) be the probability that all nodes in set A are silent at any given instant of time. According to CSMA, a node can transmit only when its neighbors are not actively receiving or transmitting. This gives us 
\begin{equation}
    \label{eqn:CSMA2}
    \frac{s_{ij}}{g_{ij}} = P(N_{i} \cup N_{j})
\end{equation} 
The network state for given $g_{ij}$ is defined by the set of nodes D that are in the transmitting state. Using steady-state probabilities and global balance equations P(A) can be given as
\begin{equation}
    \label{eqn:CSMA3}
    P(A) = \frac{SP(A^{c})}{SP(V)}
\end{equation}

where SP is the sum-of-products, $A^{c}$ is the set of nodes that are not in set A, and SP(V) is the sum-of-products of all the nodes in the network. At any given time, there will be \textit{D} independent sets containing \textit{i} nodes that can transmit concurrently in a CSMA network. SP can be calculated as follows: First, calculate the product of all the scheduled rates of all the nodes in set \textit{i}. Then, the product of all these independent sets is summed. SP can be expressed as:
\begin{equation}
    SP(A) = \sum_{D \in A} \prod_{i \in D} g_{ij}
\end{equation}
Now, substituting \ref{eqn:CSMA3} in equation \ref{eqn:CSMA2} gives,  
\begin{equation}
  \label{eqn:CSMA4}
  \frac{s_{ij}}{g_{ij}} = \frac{SP(([(N_{i} \cup N_{j})])^{c})}{SP(V)} , j \in N_{i}
\end{equation}
The above equation \ref{eqn:CSMA4} can be written for all the transmit/receive pairs in the network to form a system of linear equations that can be solved iteratively for $s_{ij}$ with given $g_{ij}$ which gives the end-to-end throughput of a multi-hop CSMA network~\cite{boorstyn1987throughput}. 
\begin{figure*}[t]
  \vspace{-0.1cm}
  \captionsetup[subfigure]{justification=Centering}
  \centering
  \begin{subfigure}{0.59\textwidth}
    \centering
    \includegraphics[width=6.5cm, height = 3.4cm]{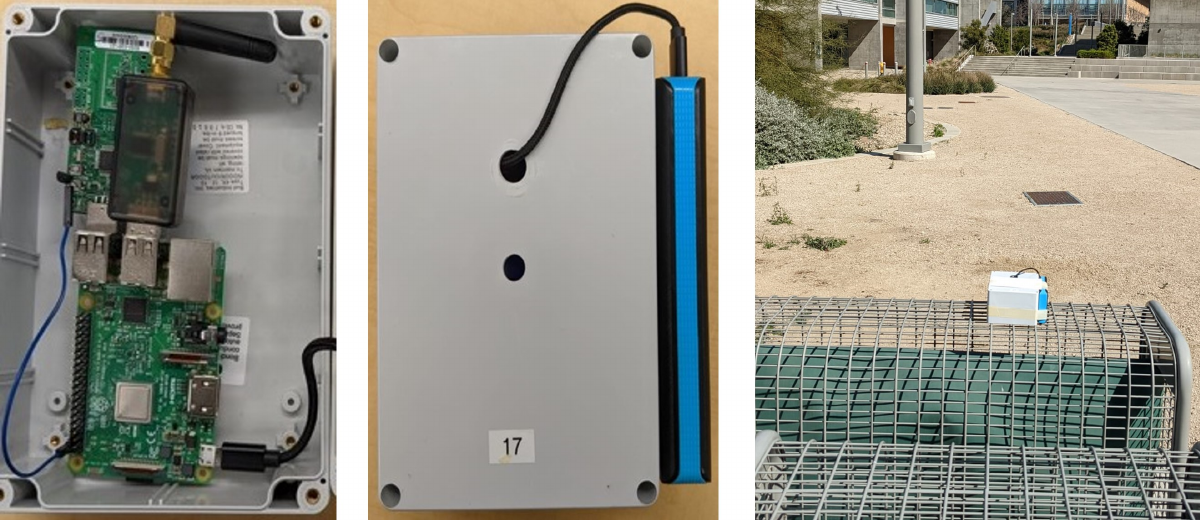}
    \caption{Multi-radio architecture node that contains both USB-based LoRa and Zigbee radios hosted by a Raspberry Pi.}
    \label{fig:Hardware}
  \end{subfigure}
  \begin{subfigure}{0.39\textwidth}
    \centering
    \includegraphics[width=5cm, height = 4cm]{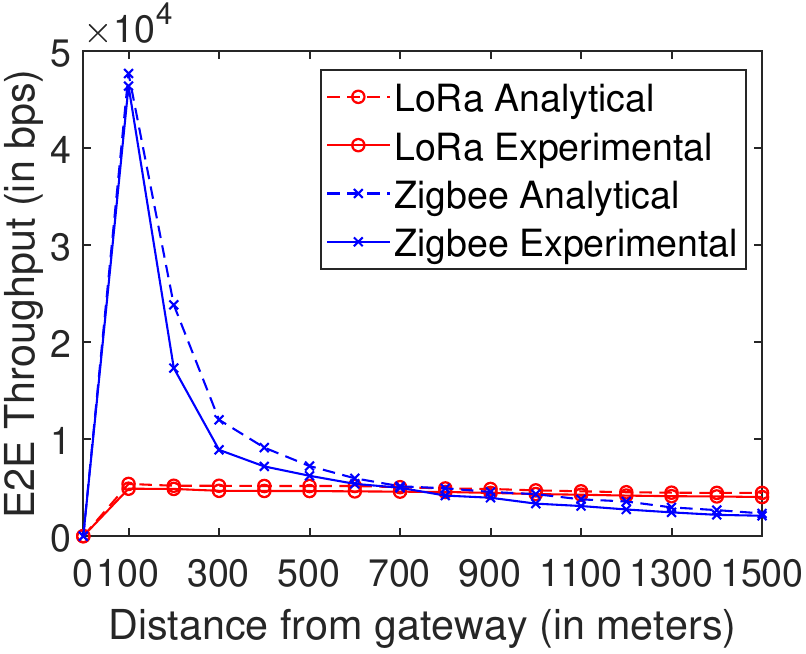}
    \caption{Average throughput at different distances}
    \label{fig:tput_dist}
  \end{subfigure}
  \caption{Hardware used and multi-hop line topology results.}
\end{figure*}


\textbf{Experimental throughput analysis}
\label{subsec:hardware}
is conducted with a multi-radio hardware node shown in Figure~\ref{fig:Hardware}. This is a Raspberry Pi 3B hosting a USB-based TelosB~\cite{telosb} Zigbee mote and a USB-based LoStik LoRa~\cite{ronoth_llc} nodes. The Raspberry Pi unit is powered by a portable external Power Bank and protected by a PVC casing. TelosB~\cite{telosb} and LoStik nodes~\cite{ronoth_llc} are programmed to transmit a 29-byte packet on receiving a command from the Raspberry Pi. During this experiment, link-level acknowledgments are disabled and CSMA is enabled in the TelosB motes. ALOHA MAC is enabled in LoRa radios. The nodes are placed in a line topology to have \textit{connected links}~\cite{baccour2012radio} to the neighboring nodes. While it is tedious to find the spots with a high packet reception ratio at longer distances, multiple iterative efforts helped to identify them.
\subsubsection{\textbf{Multi-hop result analysis}}
\label{subsec:grayregion}
 Figure \ref{fig:tput_dist} depicts the End-to-End (E2E) throughput as the function of distance. This end-to-end throughput at a specific distance is the average of one thousand packets.
The reason for a drastic difference in E2E throughput between LoRa and Zigbee at 100m is the fact that only one node is transmitting without any contention. 
A considerable drop in Zigbee throughput is seen from 100m-300m because it uses a single channel with three contenders. CSMA mechanism blocks two other links from transmitting to avoid collisions, allowing only 1 of 3 links to transmit at any given time until 300m. One could argue to use multi-channel Zigbee, but this will be detrimental for \Name\xspace (refer \S\ref{sec:limitations} for more details). 
After 300m, Zigbee's throughput steadily decreases with distance because of CSMA blocking delay and queuing delay. LoRa's throughput does not show any drastic decrease but slightly decreases with distance. This is because the LoRa signals are robust enough to pass through the mesoscale IoT application environment as the Chirp Spread Spectrum used by LoRa for modulation makes the signal highly resistant to attenuation. Also, the LoRa gateway can receive and decode eight packets concurrently. 
Figure \ref{fig:tput_dist} shows that Zigbee wins until 500m and LoRa wins after 1200m. LoRa and Zigbee achieve competitive throughput between 500m-1200m from the gateway, henceforth referred as the \textit{gray region}. If a LoRa gateway is placed at the center of the deployed environment, the \textit{gray-region} will cover an area of 3,738,495$m^2$.   

\subsection{E2E Throughput fluctuations in gray region}
\label{subsec:e2e_fluc}

\begin{figure*}[t]
  \vspace{-0.1cm}
  \captionsetup[subfigure]{justification=Centering}
  \centering
  \begin{subfigure}{0.30\textwidth}
    \includegraphics[width=4cm, height = 4cm]{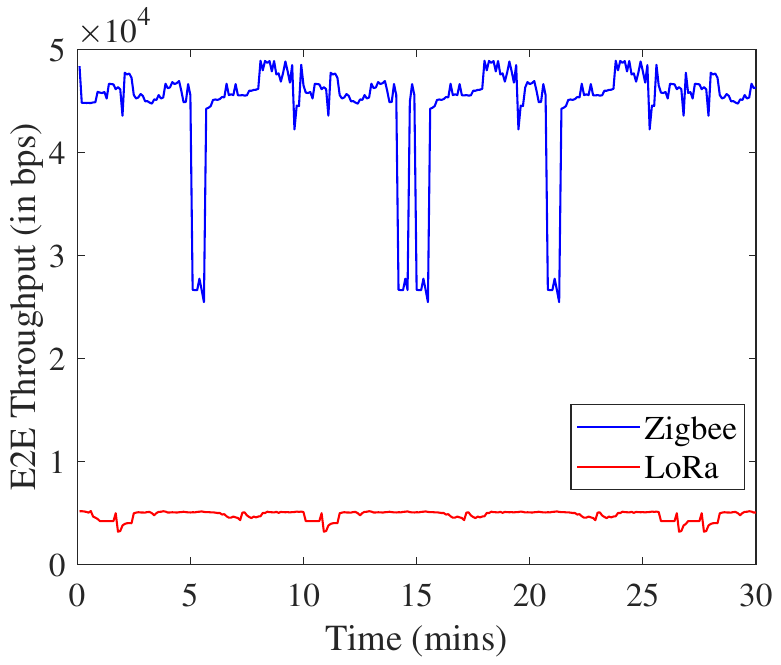}
    \caption{TF at 100m}
    \label{fig:tput_100}
  \end{subfigure}
  \begin{subfigure}{0.30\textwidth}
    \includegraphics[width=4.2cm, height = 3.8cm]{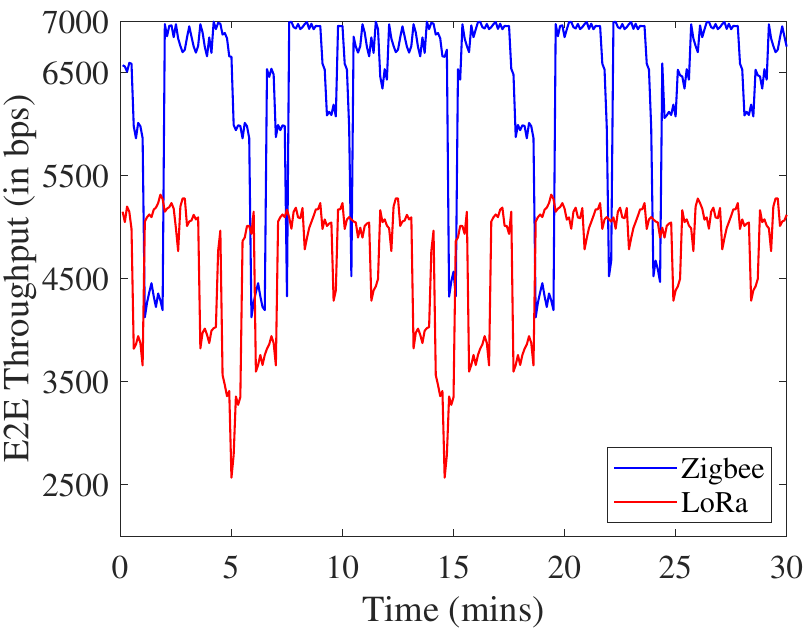}
    \caption{TF at 500m}
    \label{fig:tput_500}
  \end{subfigure}
  \begin{subfigure}{0.30\textwidth}
    \includegraphics[width=4.3cm, height = 3.8cm]{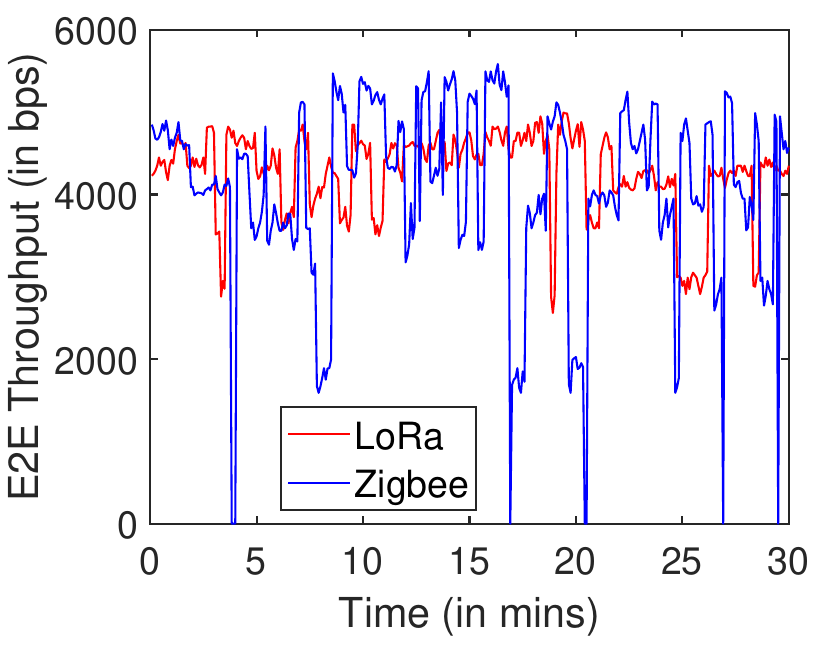}
    \caption{TF at 800m}
    \label{fig:tput_800}
  \end{subfigure}
  \bigskip

  \begin{subfigure}{0.33\textwidth}
    \includegraphics[width=\textwidth,height = 4cm, width = 4cm]{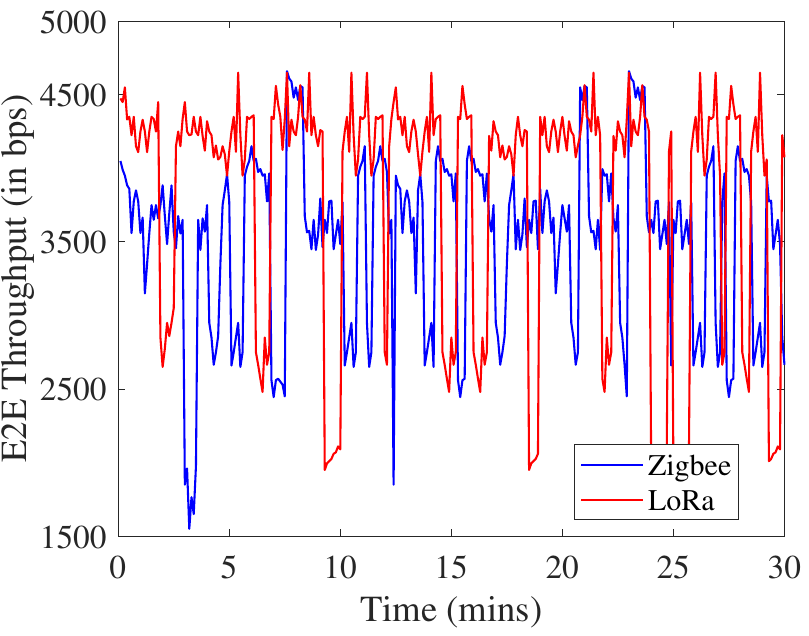}
    \caption{TF at 1200m}
    \label{fig:tput_1200}
  \end{subfigure}
  \begin{subfigure}{0.33\textwidth}
    \includegraphics[width=\textwidth,height = 4cm, width =4cm]{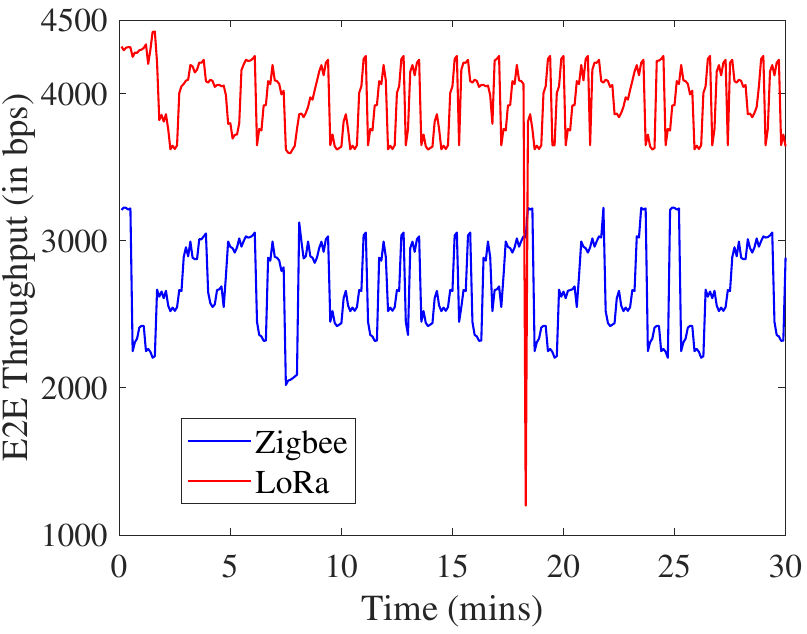}
    \caption{TF at 1400m}
    \label{fig:tput_1400}
  \end{subfigure}
  \vspace{-0.3cm}
  \caption{Throughput Fluctuations (TF) at different distances from the gateway}
  \vspace{-0.3cm}
\end{figure*}
Fig.~\ref{fig:tput_dist} shows that LoRa and Zigbee provide competitive throughput between 500m-1200m from the gateway. The end-to-end throughput at different distances from the gateway depicted in Fig.~\ref{fig:tput_dist} is the average E2E throughput of a thousand packets. This does not depict the end-to-end throughput fluctuations of different packets over time. So, we take a deeper look into the end-to-end throughput fluctuations over time at different distances from the gateway.

Figs.\ref{fig:tput_100} - \ref{fig:tput_1400} show the throughput fluctuations for 30 minutes between 100m to 1400m from the gateway. 
Zigbee wins at 100m (Fig.\ref{fig:tput_100}) and LoRa wins at 1400m (Fig.\ref{fig:tput_1400}). 
At 500m from the gateway (Fig. \ref{fig:tput_500}), Zigbee achieves higher throughput most of the time. Whenever Zigbee's throughput is falling, LoRa is able to back up Zigbee to provide better throughput. The difference in throughput of Zigbee and LoRa is considerably higher at 500m.

Fig.~\ref{fig:tput_800} shows the throughput fluctuations at 800m. Zigbee mostly wins but the throughput of Zigbee is highly fluctuating as packets are experiencing multiple hops. The average throughput of Zigbee and LoRa is very close to each other. Fig.~\ref{fig:tput_1200} shows that Zigbee experiences low E2E throughput. So, it mostly underperforms at this longer distance of 1200m from the gateway. 

The fundamentally surprising information here is that two radios with an order of magnitude difference in theoretical throughput, are achieving competitive throughput performance in the gray region.
From the above throughput fluctuations, it is evident that using a single radio IoT network leads to throughput loss even on a simple line topology. While most of the real-world IoT applications will employ the more complex mesh topology, this throughput loss will get further amplified in real-world mesh topology applications. \\

\textbf{Key observations:} 
\begin{enumerate}
    \item There are no dedicated radios for mesoscale IoT applications.
    \item Qualitative analysis identified Zigbee 2.4GHz and LoRa 915MHz as radio candidates. Quantitative multi-hop experiments showed that Zigbee and LoRa radio achieve competitive throughput from 500-1200m from the gateway, called the \textit{gray region}. 
    \item Erratic end-to-end throughput fluctuations were observed in the gray region. Predicting these throughput fluctuations will help to select a higher throughput radio at the time of transmission. 
    \item Employing radios in a multi-hop fashion, heavily reduces end-to-end throughput.
\end{enumerate}

\section{Why multi-radio for mesoscale IoT networks?}
\label{sec:motivation}
\begin{figure}[t]
  \centering
  \vspace{-0.1cm}
  \includegraphics[height=3.5cm, width=8cm]{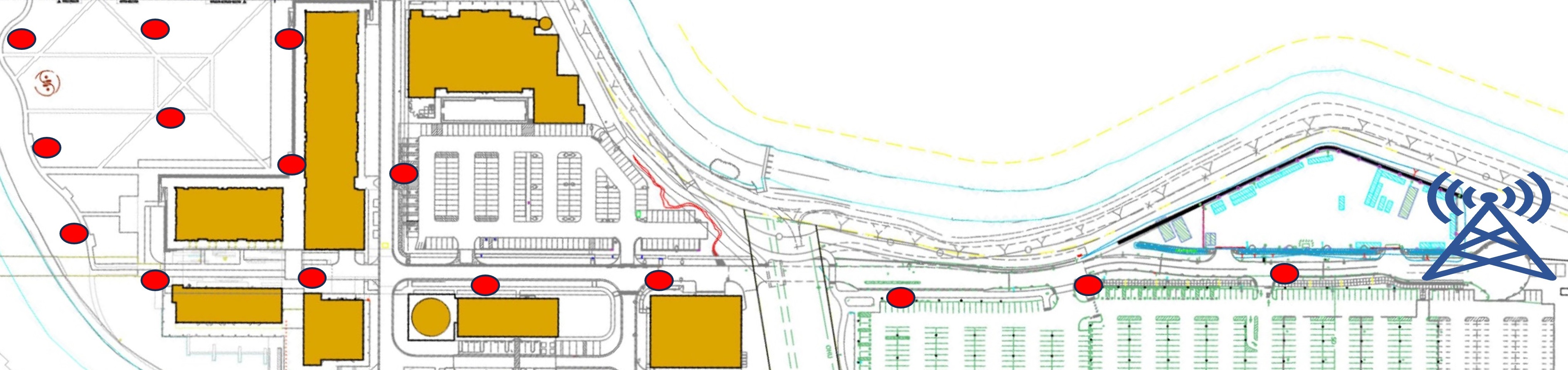}
  \caption{Location A Mesh topology - nodes populated at the gray region.}
  \label{fig:LocA-mesh-topo}  
  \vspace{-0.6cm}
\end{figure}
The above key observations made on a simple line topology suggest that Zigbee and LoRa together will provide higher throughput for mesoscale IoT applications. Line topology is seldom used in real-world applications. Hence, it will be interesting to explore the throughput fluctuations of LoRa and Zigbee in a mesh topology.  

\textbf{Experimental setup.} In this preliminary motivation experiment, nodes are populated in the gray region to form a mesh topology as shown in Figure \ref{fig:LocA-mesh-topo}. The hardware setup explained in section \ref{subsubsec:multiho_exp} is utilized. A total of one thousand 29-byte packets are transmitted by each node in the network destined for the gateway. The R Pi host commands both radios every 3 seconds to transmit a packet concurrently. The LoRa radios use ALOHA MAC. The Zigbee radios use CSMA MAC and link-level acknowledgments are enabled in TinyOS. After multiple iterative efforts, every Zigbee node is placed in a position that provides a connected link~\cite{baccour2012radio} to its immediate neighbors. 

\textbf{MAC protocol.} Although concurrent communications~\cite{ferrari2011efficient} and TDMA~\cite{fu2024comprehensive, huang2012evolution} approaches may outperform CSMA, we considered CSMA MAC since the mesoscale applications will generate different payloads at different times, that needs to be transmitted immediately. This is better suited for CSMA than TDMA/Concurrent communications. Concurrent communications requires time-synchronized nodes to leverage constructive interference. 
In mesoscale IoT applications, like P2P energy trade, two IoT nodes may not be tightly time-synchronized. If an IoT end-node wants to send an inquiry for energy trade, it has to send the message immediately without waiting for its next time slot for transmission. Hence, CSMA MAC is the right MAC choice for mesoscale IoT applications.
The Zigbee radios employ a Distance-Vector protocol~\cite{cheng1989loop} for multi-hop routing. 

\begin{figure}[t]
  \centering
  \includegraphics[width=0.3\textwidth, height = 4cm]{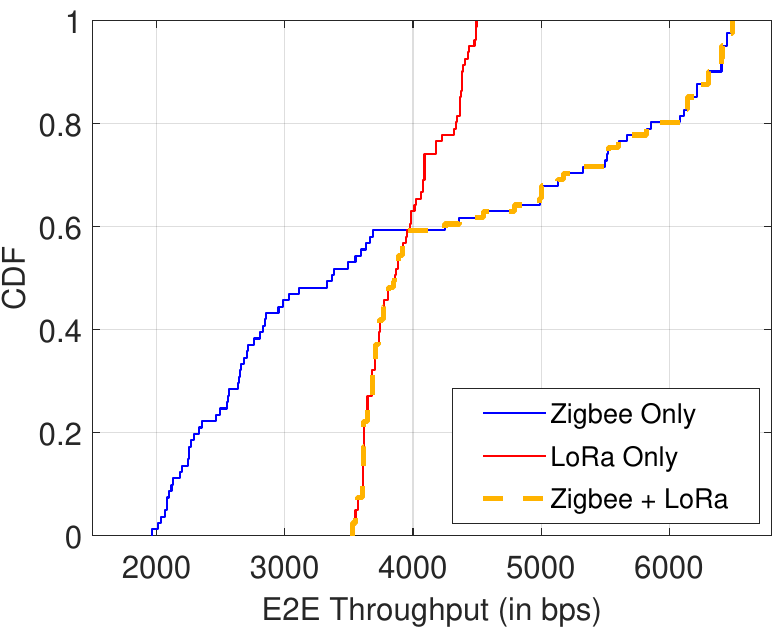}
  \vspace{-0.3cm}
  \caption{Multi-radio system with Zigbee+LoRa radios achieves higher throughput than single-radio systems.}
  \label{fig:motcdf}
\end{figure}

\textbf{Results.} The throughput achieved by both the radios of the network is plotted as a CDF in Fig.~\ref{fig:motcdf}. This figure shows that LoRa achieves higher throughput for 59\% of the transmissions and Zigbee achieves higher throughput for 41\% of the transmissions.  
The throughput is calculated as the fraction of total bits sent over the latency incurred. Throughput has an inverse relationship with latency.
According to 5G America's report~\cite{5Gamericas}, the average required latency for mesoscale IoT applications is 55ms. The average latency achieved by Zigbee-only and LoRa-only radios is 62.32 ms and 66.55 ms respectively. This is higher than the average required latency, 55ms.  

\begin{table}[t]
  \centering
  \caption{Latency of different radio systems}
  \label{tab:mot_lat_table}
  \vspace{-0.2cm}
  
  \begin{tabular}{  c  c  c c c }
    \toprule
      &\thead{Average Required \\Latency}& \thead{Zigbee+\\LoRa} &\thead{Zigbee \\ Only}&\thead{LoRa\\Only}\\
    \toprule
      &55ms& \cellcolor{green}50.62ms& \cellcolor{red}62.32ms& \cellcolor{red} 66.55ms\\
    \bottomrule
    \vspace{-0.3cm}
  \end{tabular}
\end{table}

Based on the above observations, a trace-driven simulation is conducted on the multi-radio network containing both Zigbee and LoRa radios. This trace-driven simulation mimics the performance of multi-radio networks that can choose higher throughput radio for every transmission. This is plotted as the dashed golden line in figure \ref{fig:motcdf}. This dashed golden line perfectly traces LoRa radio until the first 59\% of the transmissions and follows Zigbee radio for the next 41\% of the transmissions. 
The average latency achieved by this simulated multi-radio network maximizing for throughput, 50.62ms, falls within the bounds of the average required latency.
This shows the necessity of an intelligent multi-radio system for mesoscale IoT applications that can choose a higher throughput radio at the time of transmission.  

While it is convenient to choose a high-throughput radio based on traces, it is tedious to predict a high-throughput radio in real-world deployments. The end-to-end throughput fluctuates over time as shown in Figures \ref{fig:tput_100} - \ref{fig:tput_1400}. If an end node is able to predict these throughput fluctuations, a high-throughput radio can be predicted during transmission. 

\section{Building a machine learning model}
\label{sec:ML_build}

This section explains the process of building a Machine-Learning (ML) model for instantaneously selecting a high-throughput radio at the time of transmission. This radio selection problem is formulated as a classification problem in \S\ref{subsec:Problem_formulation}. Data is collected on three different topologies. "Location A - Line" is the Line topology described in \S~\ref{subsubsec:multiho_exp} and "Location A - Mesh" is the mesh topology described in \S~\ref{sec:motivation}. "Location B 
 - Mesh" is the mesh topology mentioned in \S~\ref{sec:eval}. 
 We develop separate ML models for each topology because our efforts for Model Retraining did not yield fruitful results since the signal attenuation, fixed, and moving obstacles are different at each location (refer \S\ref{sec:limitations} for more details). 
 
 Each topology consists of 15 end nodes and one gateway. Since our region of interest is in the \textit{gray region}, nodes are populated in this region to form a mesh topology. Data packets transmitted by the nodes in the \textit{gray region} are considered for model training and testing. During this data collection experiment, the host Raspberry Pi will command both radios to transmit a 29-byte data packet concurrently. 
 A total of 25,500 data packets were recorded. This comprehensive data set covers all the different dynamics of the deployed environment. The throughput of each transmitted data packet is recorded.
 This data set is manually labeled by a human to identify the high-throughput radio for each transmitted data packet. LoRa radios recorded the RSSI of the ACK sent by the gateway to acknowledge the previous data packet~(refer \S\ref{subsec:Problem_formulation} for more details). Zigbee radios sent beacon packets every 30ms to estimate and record local link qualities at each end node. These local link qualities were utilized to manually calculate the path quality metrics. This manually calculated path quality metric is then fed as an input to the ML models.
 
 Three widely used classification models, Support Vector Machine (SVM), Logistic Regression (LR), and CART Decision Trees (DT) are trained and tested using the SKLearn library. In addition to these three models, the CART is further optimized with the Tree Alternating Optimization (TAO) algorithm~\cite{carreira2018alternating}. The TAO-optimized CART model outperforms all the other models. 
 The TAO-optimized CART model is deployed as IF...ELSE statements in IoT end devices. This model gets the input features from path quality estimations (see \S\ref{subsec:Problem_formulation} and \S\ref{subsec:DT_est}).

\subsection{Problem formulation}
\label{subsec:Problem_formulation}

The classification model takes the input features 
E2E path quality of LoRa radios (E2E-$PQ_{LoRa}$), and the E2E path quality of Zigbee radios (E2E-$PQ_{Zigbee}$) to output a high throughput radio. The input feature vector can be expressed as:
\begin{equation}
  \label{eq:input_vector}
  Input_{i} = [E2E-PQ_{LoRa}, E2E-PQ_{Zigbee}]        
\end{equation}


The output of the machine learning model is the radio predicted to have higher instantaneous throughput. This can be expressed as:
\begin{equation}
  \label{eq:output_vector}
  Output_{i} = [Zigbee | LoRa]  
\end{equation}

\textbf{\textit{Feature selection and engineering.}} The classification problem formulated above needs E2E-$PQ_{LoRa}$, E2E-$PQ_{Zigbee}$ to predict the instantaneous high throughput radio. Traditional path-quality estimation in multi-hop Zigbee networks utilizes the link quality estimations of all the links along a path. The most common E2E path quality metrics used for multi-hop Zigbee networks are $Hop\_Number$ (HN), Packet Reception Ratio (PRR), Expected Transmission Count (ETX)~\cite{de2003high}, and Required Number of Packets (RNP)~\cite{cerpa2005temporal}. HN is the node's distance from the gateway in terms of hops. It is a discrete value ranging from [5,12] inclusively. It is obtained via the distance vector routing protocol run by Zigbee radios. The high data rate Zigbee Radio frequently transmits short beacon packets to estimate the path quality metrics. PRR is a well-known metric calculated as the ratio of the total number of packets received over the total number of packets sent by an end device. 
We calculate PRR, RNP and ETX over a window of size $\alpha$. $\alpha$=10 gave us better results in our experiments, although this may vary for different deployments.  
ETX considers both forward and backward link qualities to calculate the metric. In our case, only the forward link quality is required to estimate E2E path quality estimation from an end device toward the gateway. So, the ETX becomes $1/PRR$, making this a redundant metric in the presence of PRR. RNP has the unique characteristic of capturing the underlying distribution of packet losses~\cite{cerpa2005temporal}. So the HN, E2E PRR, and E2E RNP path quality metrics are considered for E2E-$PQ_{Zigbee}$.  These metrics are calculated with frequent beacon packets since Zigbee is a high-data-rate radio.

On the other hand, the communication channels of low data-rate LoRa radio will be clogged if frequent beacon packets are sent. Hence the RSSI of the ACK sent by the gateway for the previous data packet is considered for estimating the end-to-end path quality of the LoRa radios (E2E-$PQ_{LoRa}$).
This might be counter-intuitive because of two reasons:
\begin{enumerate}
    \item How is it possible to predict the uplink channel quality based on a downlink packet? Quail~\cite{gadre2020quick} identifies and utilizes \textit{Channel Reciprocity} between LoRa's uplink and downlink. This characteristic is leveraged to solve this issue. 

    \item How long is the uplink quality correlated?  
\end{enumerate}

\textbf{How long is the uplink quality correlated?} The inter-packet interval of data packets is 3 seconds. The time interval between the ACK corresponding to a previous data packet and the next data packet will be at least 2.5 seconds. If the metric related to a previous packet is utilized to predict the channel for the next data packet, it is not clear whether the uplink quality will be the same during this time interval. This question is answered through experiments. 

An end node transmits a 29-byte packet to the gateway every second. The RSSI value of these packets is recorded at the gateway. The autocorrelation between the RSSI values of two packets is analyzed as the Conditional Probability of two events: These two packets have (i) the same RSSI values, (ii) different RSSI values. This experiment is conducted in both free space and built environments. 

\begin{figure}[t]
  \centering
  \vspace{-0.1cm}
  \includegraphics[height=5cm,width=6cm]{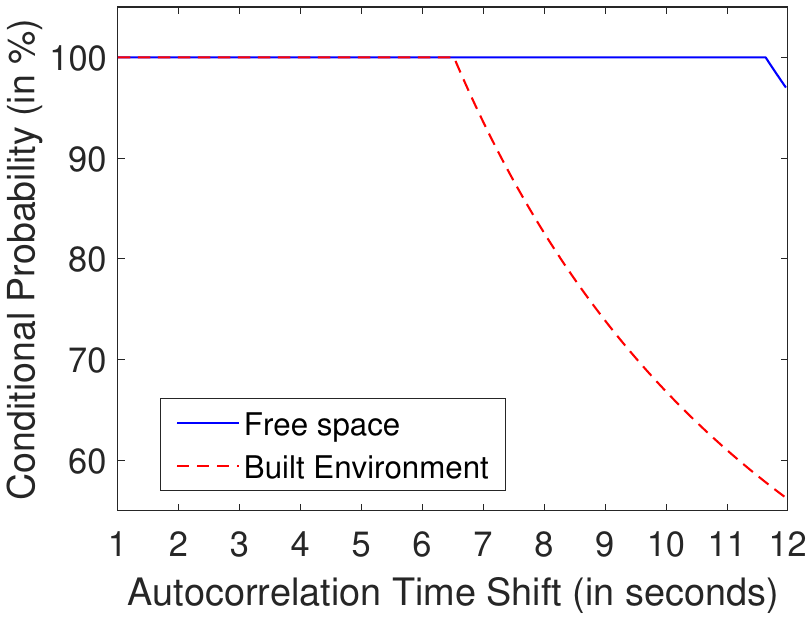}
  \vspace{-0.2cm}
  \caption{RSSI auto-correlation of LoRa links}
  \label{fig:autocorr}
  \vspace{-0.4cm}
\end{figure}

Figure \ref{fig:autocorr} shows that the channel quality is correlated for 11.6 seconds in free space and 6.5 seconds in built environments during our experiments. So, if the packet generation interval between the data packets is less than the above-mentioned duration for our deployed environment, this feature works well as the end-to-end path quality indicator for LoRa. 
If no packet is sent for 6.5 seconds, we transmit a dummy packet to preserve this correlation.

After finalizing all the input features, the input feature vector eq. \ref{eq:input_vector} becomes: 
\begin{equation}
  \label{eq:input_vector_mod}
  Input_{i} = [HN,E2E\_RSSI_{LoRa}, E2E\_PRR_{Zigbee}, E2E\_RNP_{Zigbee}]        
\end{equation}


\vspace{-0.3cm}
\subsection{Prediction methods and results}
\label{subsec:ML_models}
\begin{table*}[ht]
\centering
\tiny
\vspace{-0.1cm}
\caption{Training and Testing accuracy of different ML models and optimizations for all the topologies}
\label{tab:model_pred_results}
\vspace{-0.2cm}
\resizebox{\textwidth}{!}{%
\begin{tabular}{|c|cccc|cc|}
\hline
\multirow{3}{*}{\begin{tabular}[c]{@{}c@{}}ML\\Models \\and\\ Optimizations\end{tabular}} &
  \multicolumn{4}{c|}{Location A} &
  \multicolumn{2}{c|}{Location B} \\ \cline{2-7} 
 &
  \multicolumn{2}{c|}{Line} &
  \multicolumn{2}{c|}{Mesh} &
  \multicolumn{2}{c|}{Mesh} \\ \cline{2-7} 
 &
  \multicolumn{1}{c|}{\begin{tabular}[c]{@{}c@{}}Training Accuracy\\  (in \%)\end{tabular}} &
  \multicolumn{1}{c|}{\begin{tabular}[c]{@{}c@{}}Testing Accuracy\\ (in \%)\end{tabular}} &
  \multicolumn{1}{c|}{\begin{tabular}[c]{@{}c@{}}Training Accuracy    \\ (in \%)\end{tabular}} &
  \begin{tabular}[c]{@{}c@{}}Testing Accuracy\\ (in \%)\end{tabular} &
  \multicolumn{1}{c|}{\begin{tabular}[c]{@{}c@{}}Training Accuracy\\  (in \%)\end{tabular}} &
  \begin{tabular}[c]{@{}c@{}}Testing Accuracy\\ (in \%)\end{tabular} \\ \hline
SVM &
  \multicolumn{1}{c|}{83.59$\pm$1.60} &
  \multicolumn{1}{c|}{83.37$\pm$2.51} &
  \multicolumn{1}{c|}{82.15$\pm$0.89} &
  80.87$\pm$3.50 &
  \multicolumn{1}{c|}{78.31$\pm$1.90} &
  76.62$\pm$4.65 \\ \hline
LR &
  \multicolumn{1}{c|}{83.65$\pm$1.01} &
  \multicolumn{1}{c|}{83.87$\pm$3.566} &
  \multicolumn{1}{c|}{82.59$\pm$0.83} &
  81.25$\pm$3.46 &
  \multicolumn{1}{c|}{81.21$\pm$0.57} &
  80.50$\pm$2.21 \\ \hline
CART &
  \multicolumn{1}{c|}{\textbf{93.00$\pm$0.49}} &
  \multicolumn{1}{c|}{88.00$\pm$3.40} &
  \multicolumn{1}{c|}{\textbf{92.53$\pm$0.52}} &
  83.75$\pm$1.97 &
  \multicolumn{1}{c|}{90.56$\pm$0.25} &
  82.37$\pm$2.94 \\ \hline
TAO-CART &
  \multicolumn{1}{c|}{93.00$\pm$0.56} &
  \multicolumn{1}{c|}{\textbf{88.00$\pm$2.33}} &
  \multicolumn{1}{c|}{89.375$\pm$0.26} &
  \textbf{85.625$\pm$4.12} &
  \multicolumn{1}{c|}{\textbf{94.71$\pm$0.38}} &
  \textbf{93.87$\pm$1.69} \\ \hline
\end{tabular}}
\vspace{-0.4cm}
\end{table*}
The three widely used classification models, namely Logistic Regression (LR), Support Vector Machine (SVM), and CART Decision Trees (CART) were trained using the trace-driven data set obtained from large-scale real-world experiments from different topologies. 
An ML model is built offline for each location. 
\Name\xspace will deploy the chosen ML model in all the nodes in that specific topology.
Their training and testing accuracy are averaged over 5-fold cross-validation on a data set based on real-world experiments. 
The testing and training accuracy tabulated in Table \ref{tab:model_pred_results} shows that the TAO-optimized CART (TAO-CART) is better for deployment than the other widely used models. 

CART Decision Tree Classifier~\cite{praagman1985classification}
is a classical and one of the most popular algorithms to train a DT. 
The TAO algorithm~\cite{carreira2018alternating}, takes an initial tree, either generated randomly or induced by traditional algorithms (e.g. CART), and optimizes it jointly over the parameters of all the nodes in the tree. TAO works in alternating optimization fashion by cycling over different depths of a tree. At a given depth, TAO optimizes all nodes in that specific depth in parallel while guaranteeing a monotonic decrease of the desired objective function, such as misclassification errors. This fact backed with empirical evaluations~\cite{ZharmagCarreir20a} makes TAO an attractive algorithm.

The training and testing accuracy of the widely used classification models, namely SVM, LR, and CART, is tabulated in Table \ref{tab:model_pred_results}. From these results, it is clear that CART can achieve higher accuracy than SVM and LR. Compared with CART, TAO-CART achieves similar accuracy for the simple "Location A - Line topology" which are seldom used in real-world applications, and higher accuracy than CART for the "Location B-Mesh" topology. Also, the difference between the Training and Testing accuracy of TAO-CART is comparably smaller than that of CART for all the topologies. This means that TAO-CART optimization highly generalizes the model for unseen data.
\begin{figure}[t]
  \centering
  \includegraphics[width=0.4\columnwidth, height=5cm]{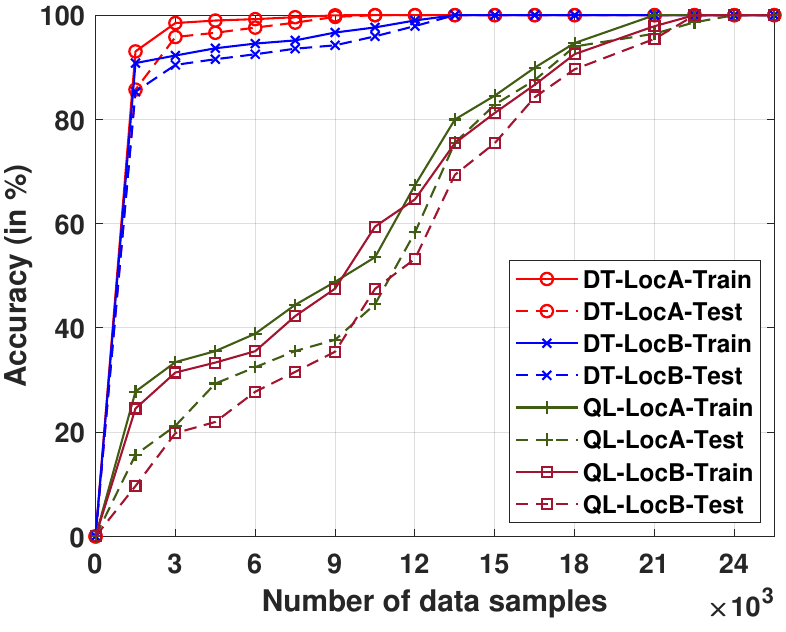}
  \vspace{-0.3cm}
  \caption{Training data requirements}
  \label{fig:train_data_req}
  \vspace{-0.7cm}
\end{figure}

\textbf{Training data requirements.} A major concern of a data-driven approach is the data required to train a model. Fig.~\ref{fig:train_data_req} plots the accuracy of CART Decision Tree (DT) model  and Q-Learning~\cite{gummeson2009adaptive} (QL) as the function of required training data samples. In Fig.~\ref{fig:train_data_req}, we can see that the DT can quickly reach over 93\% (Location A) and 90\% (Location B) training accuracy with 1500 samples and show a gradual increase to 100\% with 9000 (Location A) and 13500 (Location B) samples. Since there is only a gradual increase in accuracy after 1500 samples, we trade off accuracy to reduce the data collection effort.  This trade-off reduces the deployment effort by 88.99\%. So, all the models listed in Table \ref{tab:model_pred_results} were trained with 1200 samples and tested on 300 samples. Parametric details of this QL are given in \S\ref{sec:eval}. A well-known issue of QL (model-free reinforcement learning) is the humongous training data requirement~\cite{ding2020mb2c}. 
From Fig.~\ref{fig:train_data_req}, it is evident that using a DT model achieves significantly higher accuracy than QL with 1500 data samples. It should also be noted that QL needs an order of magnitude higher data samples to achieve the accuracy of DT. We optimized DT with TAO~\cite{carreira2018alternating}.

\section{Realizing TAO-CART on {IoT} devices}
\label{subsec:implementation}

The TAO-CART radio selector is found to achieve higher accuracy. However, realizing and deploying TAO-CART on an IoT end device entails the following challenges:

(i) \textit{Instantaneous path quality estimations.} The TAO-CART radio selector needs instantaneous end-to-end path quality as an input to accurately predict the high-throughput radio. The models trained and tested in the previous section used a trace-driven data set, whose multi-hop end-to-end path qualities of Zigbee radios were manually calculated. In real-world deployments, the local link qualities should be propagated throughout the network for path quality estimation. However, this network-wide multi-hop link quality propagation may not be instantaneous at all distances from the gateway due to the chaotic link quality variations, queuing and propagation delays. We solve this challenge by developing a DT-based instantaneous path quality estimator (refer \S\ref{subsec:DT_est}).

(ii) \textit{Model size and Inference Latency.} Deploying TAO-CART with a minimal memory footprint on resource-constrained IoT devices is crucial. TAO-CART outputs a tree-like structure for predicting the high-throughput radio. This tree-like model, converted as IF...ELSE statements, take 36KB of memory on disk. In Raspberry Pi, this model takes 0.008ms for selecting the high-throughput radio. This deployment is highly feasible on resource-constrained IoT end devices because of the lower time and space complexity, although we are certain that this has a larger margin for optimization.
\vspace{-0.3cm}
\subsection{ Traditional path quality estimations are not instantaneous}
\label{sec:trad_not_inst}
An intuitive idea to identify end-to-end throughput fluctuations is to utilize the end-to-end Path Quality (PQ) metrics. Since LoRa is a \textit{single-hop} network, E2E PQ can be easily estimated.  However, traditional PQ estimation for a \textit{multi-hop} ZigBee network may not be instantaneous at all distances from the gateway.

\begin{figure}[t]
  \centering
  \includegraphics[height=3.5cm, width=0.55\linewidth]{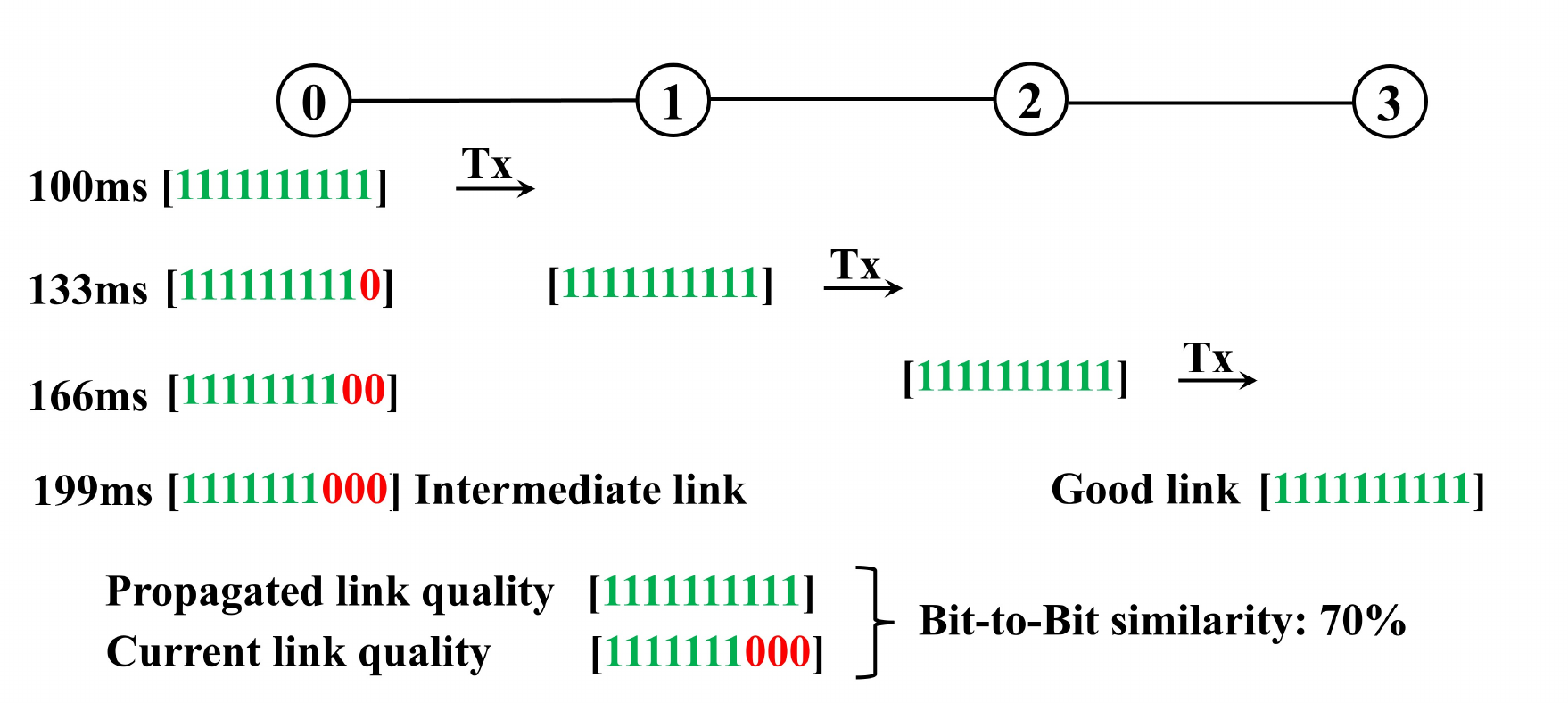}
  \caption{Traditional path quality propagation is not instantaneous at all distances from the gateway.}
  \label{fig:sprinkler_depiction}
\end{figure}

Traditional PQ estimations for a \textit{multi-hop} ZigBee network need the link quality estimation of all the links along the path. In section \ref{subsec:ML_models}, TAO-CART was trained with a trace-driven data set, whose path qualities were manually calculated based on the recorded local link qualities. In practice, the qualities of all the links along the path should be propagated throughout the network so that they can be used to estimate the path quality. The packet in which these link qualities are propagated will experience delays at different hops due to (i) wireless link quality variations and (ii) packet processing delay. Packet processing delay includes read/write delays to append link quality sequences of the intermediate nodes, link-layer Acknowledgment delay, CSMA delay, propagation and queuing delays along a multi-hop network. So, the propagated local link quality may expire when it reaches a node estimating the metric.


The problem of traditional PQ estimation is depicted in Fig.~\ref{fig:sprinkler_depiction}. In this figure, Node 1 sends a beacon packet to Node 0 every 30ms. Node 0 stores the reception and loss of the beacons as a (1/0) binary bit-sequence respectively. It is identified through experiments that propagating a packet containing this link quality sequence takes 33ms on average for a single hop transmission accounting for all the above-mentioned delays while the network has fully functional control and data planes. Figure~\ref{fig:sprinkler_depiction} depicts that the link quality sequence of link 1-0 propagated by Node 0 becomes invalid when it reaches Node 3. The bit-to-bit similarity of the link sequence of Node 0 at 199ms is not the same as propagated by Node 0. The bit-to-bit similarity reduces to 70\% at the third hop. This problem will further be amplified if all the nodes along the path append their link quality sequence to this packet for network-wide advertisement. Since the gray region is 500m-1200m from the gateway, estimating path quality with expired link quality sequences will not be instantaneous, and this cannot act as an indicator of E2E throughput fluctuations.   


\subsection{DT-based path quality estimation}
\label{subsec:DT_est}
DT-based path quality estimation is developed to mitigate the problem of conventional path quality estimation for multi-hop networks. While the conventional path quality is calculated with Link Quality(LQ) information from all the links encompassed by the end-to-end path, DT-based path quality estimation requires LQ information only from a portion of the entire end-to-end path to compute the PQ metrics. This will highly reduce the delay incurred to propagate LQ information in a multi-hop Zigbee network. DT-based path quality estimation surfaces from the two important observations listed below: (i) The link quality of each link in the path may independently change based on the deployed environment and (ii) Each end node runs \Name\xspace to select a radio for transmitting the packet. A path from an end node to the gateway consists of multiple links. While it is intuitive to understand that LQ information from a portion of the entire end-to-end path is enough to compute PQ metrics, the challenge here is to identify the required path length, $RP_n$, so that the TAO-CART radio selector can accurately predict the high-throughput radio. For example, say a path from the end node to the gateway consists of 10 links. Traditional PQ estimation will use LQ information from all these 10 links to compute PQ metrics, whereas DT-based PQ estimation requires LQ information only from $RP_n$ (<10) links to compute PQ metrics. The problem here is to define $RP_n$. We address this problem by training and testing decision trees with PQ metrics computed from different partial path lengths ($RP_n$) to understand the prediction accuracy of Decision Trees (DT).
\begin{figure*}[t]
  \vspace{-0.1cm}
  \captionsetup[subfigure]{justification=Centering}
  \centering
  \begin{subfigure}{0.32\textwidth}
    \includegraphics[width=\textwidth, height = 3.7cm]{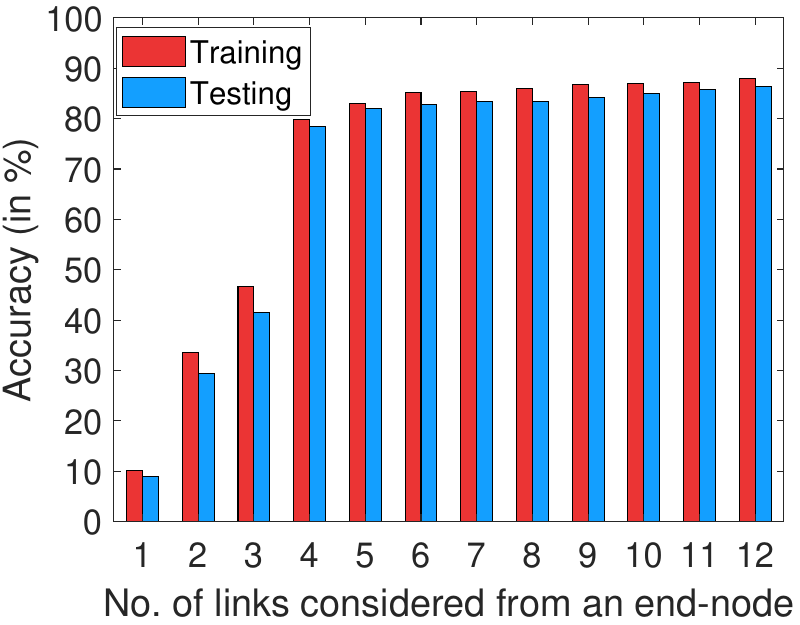}
    \caption{$RP_n$=4 - "Location A - mesh"}
    \label{fig:gl_locA}
  \end{subfigure}
  \begin{subfigure}{0.32\textwidth}
    \includegraphics[width=\textwidth, height = 3.7cm]{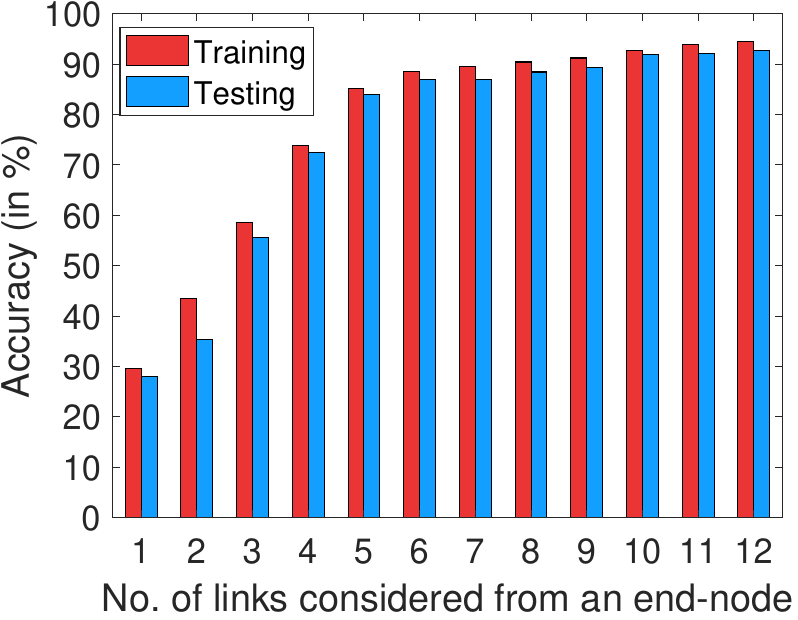}
    \caption{$RP_n$=5 - "Location B - mesh"}
    \label{fig:gl_locB}
  \end{subfigure}
  \begin{subfigure}{0.32\textwidth}
    \includegraphics[width=\textwidth, height = 3.7cm]{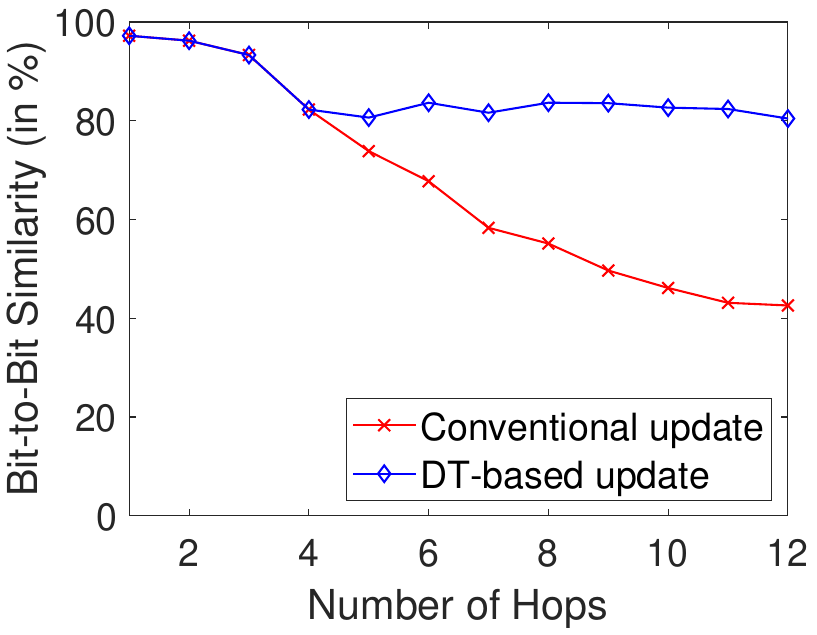}
    \caption{DT-based PQ estimation}
    \label{fig:DT-sim}
  \end{subfigure}
  \caption{DT-based PQ estimation. In figures \ref{fig:gl_locA}, \ref{fig:gl_locB}, the x-axis denotes the number of links considered to calculate PQ metrics.}
\end{figure*}

A decision tree takes the end-to-end path quality metric of both the radios as input and chooses one radio for transmitting the packet. The following test is conducted to identify this Required Path length $RP_n$: The path quality metrics are calculated with LQ information from the different number of links, from an end node towards the gateway. The training and testing accuracy of the models using LQ information from the different number of links along the path for mesh topologies at Locations A and B are depicted in Figures.~\ref{fig:gl_locA} and \ref{fig:gl_locB} respectively. From these figures, it is inferred that there is a very gradual increase in accuracy after the fourth and fifth hops for the "LocationA-mesh" and "LocationB-mesh" topologies respectively. These decision trees are further optimized by the TAO algorithm. The above values are set to $RP_n$ for our large-scale experiments.  Figure~\ref{fig:DT-sim} shows that the bit-to-bit similarity of conventional PQ estimation decreases with increase in hops while the bit-to-bit similarity of DT-based PQ estimation averages to 81\%. 

\section{Large-scale experimental results}
\label{sec:eval}
\begin{figure}
  \centering
  \includegraphics[height=3.5cm, width=8cm]{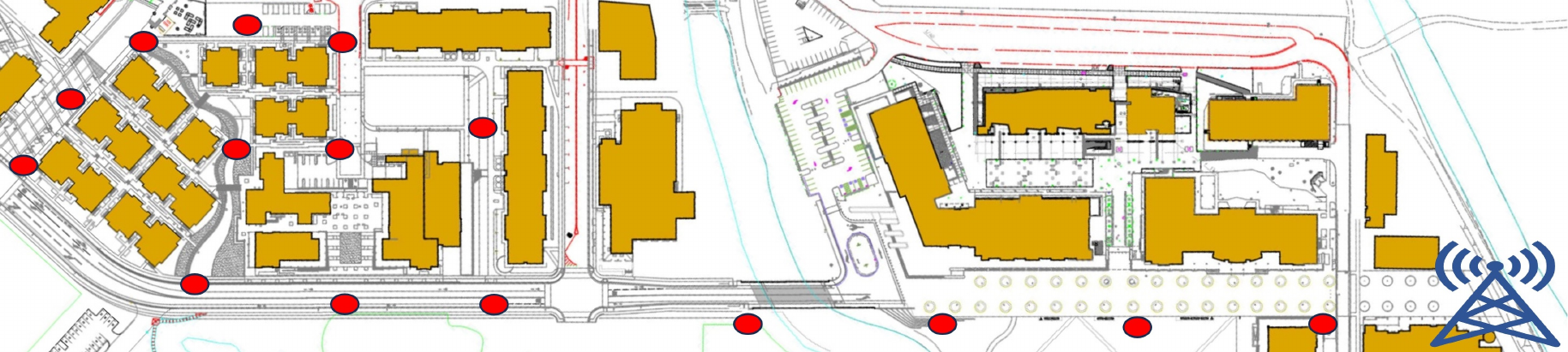}
  \caption{Location B Mesh topology - nodes populated at the gray region.}
  \label{fig:LocB-mesh-topo}
\end{figure}

\Name\xspace is evaluated through large-scale real-world, mesh topology experiments conducted on our campus. Two mesh topologies are set up in two different locations as shown in Figs.~\ref{fig:LocA-mesh-topo} and \ref{fig:LocB-mesh-topo}. 
These topologies are deployed at locations with complex environments with different building materials and heavy human influx.
Each node uses the hardware setup explained in \S\ref{subsec:hardware}. 
A total of 10,400 data packets were transmitted at both locations for this evaluation.
\begin{table}[t]
  \centering
  \caption{Thresholds identified to achieve higher throughput in the gray-region based on Lymberopoulos et al.~\cite{lymberopoulos2008towards}}
  \label{tab:threshold}
  \begin{tabular}{cccl}
    \toprule
      & 500-700m     & 700-1000m    & 1000-1200m\\  
    \toprule
    Zigbee RR         & $\geq$77\%         & $\geq$80\%         & $\geq$83\%\\
    LoRa RSSI         & $\geq$-72 dBm & $\geq$-71 dBm & $\geq$-71 dBm \\
    Fallback radio    & Zigbee       & LoRa         & LoRa\\
    \bottomrule
    \vspace{-0.2cm}
  \end{tabular}
  \vspace{-0.7cm}
\end{table}

\textbf{Benchmarks.} The performance of \Name\xspace is compared with the single radio systems formed by (i) Zigbee-only and (ii) LoRa-only radios. 
\Name\xspace is also compared with two Multi-radio systems: (i) Q-learning-based radio selector~\cite{gummeson2009adaptive} and (ii) Threshold-based radio selector~\cite{lymberopoulos2008towards}, optimizing for energy efficiency. We made our best effort to adopt these works for throughput optimization.  
\textit{(i) The threshold-based algorithm~\cite{lymberopoulos2008towards}} is based on the break-even points identified through experiments similar to Fig.~\ref{fig:tput_dist}. The thresholds are set based on a node's distance from the gateway and the instantaneous end-to-end path quality of the radios. End-to-end reception ratio and RSSI are used as the instantaneous path-quality indicator for the multi-hop Zigbee and the single-hop LoRa radios respectively. These thresholds are identified from the traces obtained from real-world experiments. This threshold-based algorithm divides the gray region into three sub-regions 500-700m, 700-1000m, and 1000-1200m based on grouping similar indicators achieving higher throughput. Thresholds are set based on these sub-regions for a fair comparison. Thresholds tabulated in Table \ref{tab:threshold} are identified via experiments to achieve higher throughput in each sub-region. A  radio having end-to-end reception ratio greater than 77\%, 80\%, and 83\% achieves higher throughput in the sub-regions 500-700m, 700-1000m, and 1000-1200m respectively while LoRa's RSSI $\geq$-72 dBm, $\geq$-71 dBm and $\geq$-71 dBm achieve higher throughput in the sub-regions 500-700m, 700-1000m, and 1000-1200m respectively. So, in the case of one radio performing better than the other, it will be indicated by the thresholds, eventually selected for transmitting the packet. From the experimental traces, the fall-back radio is identified to achieve better throughput if either both radios fall inside or outside the defined threshold region.

(ii) Q-learning-based radio selector~\cite{gummeson2009adaptive} is trained with 1500 samples since the TAO-CART model of \Name\xspace is also trained with 1500 samples. Comparing with the TAO-CART model of \Name\xspace, Q-Learning needs an order of magnitude higher data to converge (refer Fig.~\ref{fig:train_data_req}). We still compare the performance of \Name\xspace and Q-learning-based radio selector~\cite{gummeson2009adaptive} to understand the overall throughput performance of the models trained with the same amount of training data.
We follow the same strategy devised by Gummeson et al.~\cite{gummeson2009adaptive} to set Q-learning model parameters except the changes described below. They optimize for lower energy consumption. Since energy is a cost, not a reward, they consider their reward as $"-energy"$. In our case, we optimize for higher throughput, so our reward is $throughput$. Gummeson et al.~\cite{gummeson2009adaptive} \textit{initially} has two states, one for each radio. Since the transmit power level is \textit{one of the factors} directly affecting energy consumption, this two-state model can be expanded to an n-state model where each state represents a radio at a particular transmit power level. For example, four states are required for two radios, each with two transmit power levels. They also identify that an increase in the number of states increases exploration overhead or decreases exploration frequency. They solve this issue by considering only three states at a time. 

In our case, throughput/latency is directly affected by the link/path quality. We represent the link/path quality using the input feature vector described in \S\ref{subsec:Problem_formulation}. Unlike the countable/controllable transmit power levels of Gummeson et al. \cite{gummeson2009adaptive}, the link/path quality states are numerous and uncontrollable. This significantly increases the number of states leading to an increase in exploration overhead. We optimized the model's hyper-parameters through grid search.
Also, the sophisticated RL-based radio switching protocol of Gummeson et al.~\cite{gummeson2009adaptive} suffers from the below-described problems: (i) During data transfer between two radios in the communication range, the radio switching protocol, designed for energy efficiency, does a three-way handshake to switch radios. This incurs additional latency which will heavily degrade the throughput. (ii) A well-known issue of model-free RL is that it requires heavy training data to converge to an acceptable performance~\cite{ding2020mb2c} and the amount of data samples used by Gummeson et al. \cite{gummeson2009adaptive} for training is obscure. We compared the training data requirements of Q-learning and TAO-CART Decision Tree models in \S\ref{subsec:ML_models}.
\begin{figure*}[t]
  \captionsetup[subfigure]{justification=Centering}
  \centering
    \begin{subfigure}{0.49\columnwidth}
    \includegraphics[ height=6cm]{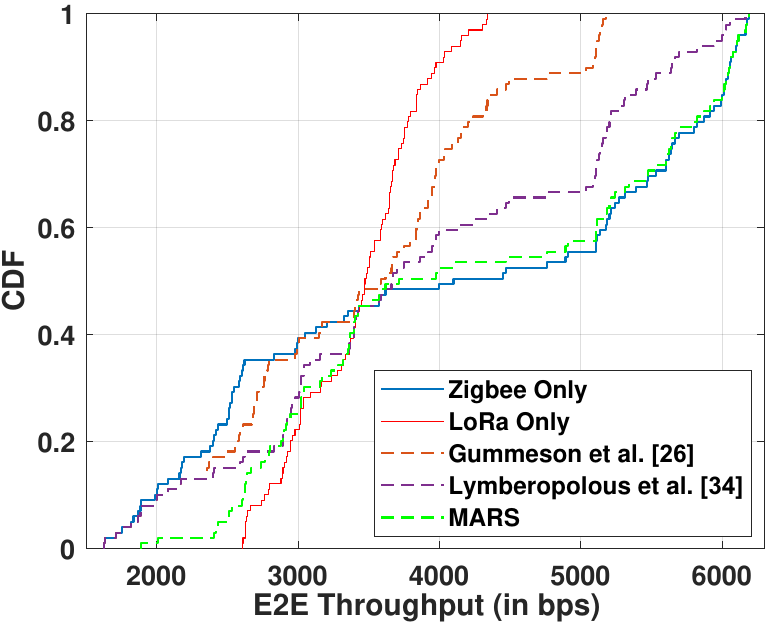}
    \caption{Throughput - Location A}
    \label{fig:locA-cdf}
  \end{subfigure}
  \begin{subfigure}{0.49\columnwidth}
    \includegraphics[ height=6cm]{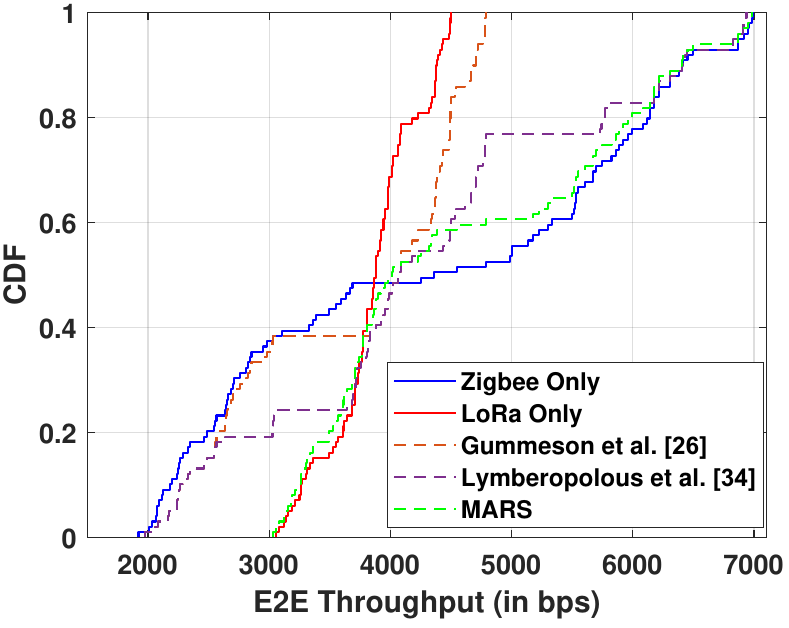}
    \caption{Throughput - Location B}
    \label{fig:locB-cdf}
  \end{subfigure}
  \caption{Throughput gain of \Name}
  \label{fig:throughput_gain}
\end{figure*}

One can argue that our design of Q-Learning for throughput maximization can still be optimized, although it will be incremental. Our Q-Learning model provides the best results after multiple iterative efforts. Moreover, our ability to optimize is also restricted in order to respect the originality of their work. It is to be noted that the above-mentioned works optimize for energy consumption. In addition, Gummeson et al.\cite{gummeson2009adaptive} use custom hardware. The performance of Gummeson et al.~\cite{gummeson2009adaptive} without their custom hardware is obscure. However, we made our best effort to adopt Gummeson et al. \cite{gummeson2009adaptive} for throughput optimization without custom hardware. 

\subsection{Performance evaluation}

The throughput gain of \Name\xspace is shown in Figure \ref{fig:throughput_gain}. The Q-learning-based radio selector~\cite{gummeson2009adaptive} achieves the least throughput gain of all the Multi-radio systems because of the three-way handshake and limited training data.

The threshold-based radio selector~\cite{lymberopoulos2008towards} fails to identify LoRa as a high throughput radio for 19\% of the transmissions. Also, the threshold-based radio selector converges towards the high-throughput Zigbee only after 60\% of the transmissions. This is because the threshold-based radio selector is not able to identify the high-throughput radio when both radios fall inside the threshold region. The identified fallback radios do not achieve higher throughput all the time because of erratic link quality variations. 

\Name\xspace closely follows the high-throughput radio as it instantaneously identifies the throughput fluctuations. In this location, the threshold-based radio selector algorithm achieves an average throughput gain of 18.79\% and 15.31\% than Zigbee-only and LoRa-only networks respectively. Whereas, \Name\xspace achieves an average throughput gain of 55.93\%, 57.22\%, and 36.32\% than Zigbee-only, LoRa-only, and Threshold-based radio selector networks respectively.  

Fig.~\ref{fig:locB-cdf} shows similar trends in Location B.
In this location, all three multi-radio systems tend to cross the solid red line unlike Fig.~\ref{fig:motcdf}, where the optimal performance was obtained offline through a trace-based evaluation. Hence, the simulated optimal multi-radio performance has to be chosen from any one of the available throughputs making the golden dashed lines of Fig.~\ref{fig:motcdf} to stay within the solid lines. The performance evaluations were conducted in real-world deployments. The channel conditions may not be identical when each system is evaluated. This change in channel conditions led to a small difference in throughput making the multi-radio systems cross the solid line.
In this location, the threshold-based radio selector algorithm achieves an average throughput gain of 16.06\%, and 19.57\% than Zigbee-only and LoRa-only networks respectively. \Name\xspace achieves an average throughput gain of 53.77\%, 58.34\%, and 32.49\% than Zigbee-only, LoRa-only, and Threshold-based multi-radio networks respectively.
\begin{figure}
  \centering
  \includegraphics[width=0.8\textwidth, height=8cm]{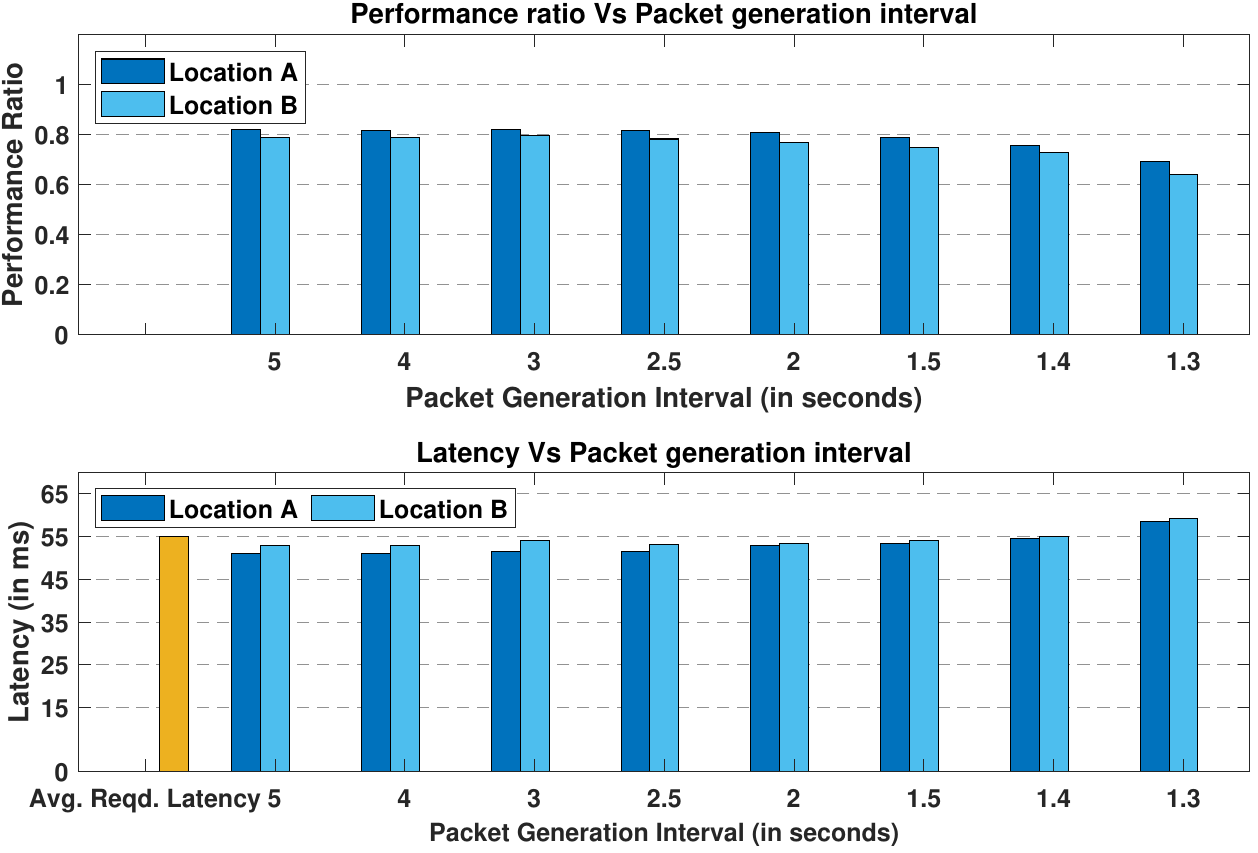}
  \vspace{-0.3cm}
  \caption{MARS on different packet generation intervals.}
  \label{fig:pack-break}
  \vspace{-0.5cm}
\end{figure}

\textbf{\Name\xspace on different packet generation intervals.} The performance ratio is calculated as the ratio of the average per-packet throughput of \Name\xspace over the average per-packet throughput of the best radio. The Performance ratio for 

different packet generation intervals is plotted in Fig.~\ref{fig:pack-break}.
We observed a slight decrease in the performance ratio from packet generation interval 5s - 1.5s and a significant decrease in performance ratio for intervals 1.4s and 1.3s. This directly reflects on the average latency. \Name\xspace is able to achieve the required latency for intervals 5-1.4s, while \Name\xspace fails to achieve the required latency for interval 1.3s. 
This degradation in \Name\xspace's performance is due to the increase in packet generation rate leading to queuing delays in the network. This increase in queuing delay affects the beacon packets used for local link quality and path quality estimations. This directly affects the required fine-grained link/path quality estimations. 

\begin{figure}
  \centering
  \includegraphics[width=0.5\textwidth, height=6cm]{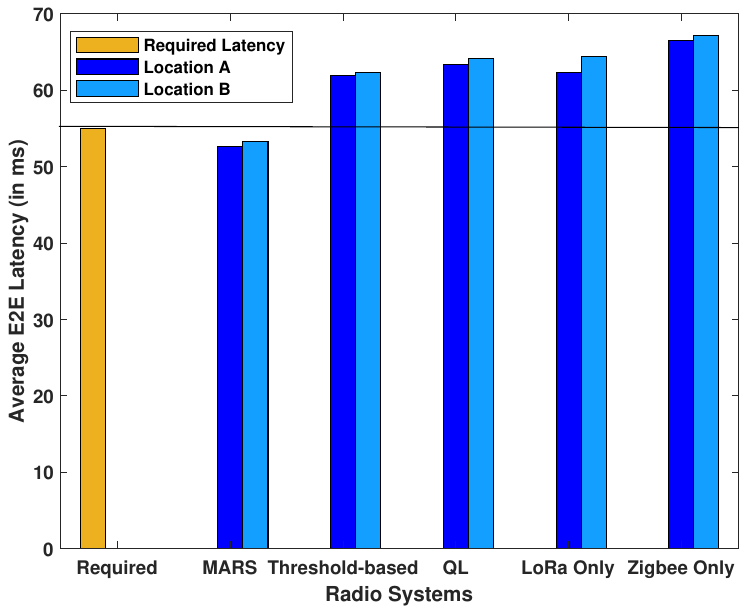}
  \caption{MARS achieves required average latency}
  \label{fig:lat-req-tao}
  \vspace{-0.4cm}
\end{figure}

\textbf{Average required latency for mesoscale IoT.} According to 5G America's report~\cite{5Gamericas}, the average required latency for mesoscale IoT applications is 55ms. Fig.~\ref{fig:lat-req-tao} shows the average latency achieved by different radio systems at two different locations. LoRa-only and Zigbee-only networks are not able to achieve the required latency in the gray region. \Name\xspace is able to achieve an average latency of 51.48ms and 53.98ms in Locations A and B respectively. \Name\xspace achieves the goal while the other radio systems fail.

\section{Discussion} 
\label{sec:limitations}

\textbf{Energy Overhead:} \Name\xspace relies on the neighbor discovery beacon packets sent every 30ms for local link quality estimation. The reception/loss of a neighbor discovery beacon, stored as a bit array, is appended to the routing table when the distance vector protocol periodically broadcasts the routing table. We piggyback on neighbor discovery beacons for local link quality estimation. This optimizes both energy consumption and queuing delays. However, appending this bit array to routing table broadcast packets leads to transmission overhead. On average, the energy toll of this additional transmission overhead per data packet is 13.60\% of the energy consumed for a data packet transmission. 
If the neighbor discovery beacons use Trickle timer~\cite{levis2004trickle}, additional beacon packets would be needed for channel estimation, significantly increasing energy consumption.
In real-world mesoscale IoT applications like P2P energy trading of smart meters in the smart neighborhood and Industrial/Factory automation, the IoT radio utilized here will be a small component of a larger device. Unlike the typical IoT nodes powered by a smaller battery, these devices are either grid-powered \cite{yang2024rateless, nagai2024improve} or have a large battery reserve to support the entire device \cite{mitsubishielectricSmartMeter, mitsubishielectricLargecapacityBattery, mitsubishielectricPowerSystems} making this energy overhead tolerable for these applications. 

\textbf{Model Retraining:} 
We evaluated \Name\xspace at different seasons of the year. The above results were obtained by deploying the ML model that was trained with data collected in the Summer and it was evaluated during the same season. Several factors affect the dynamicity of link and path qualities that makes the input feature vector of the ML model. For example, the node placement, the impact of temperature and pressure variations, and the effects of fleeting reflectors. Through experiments, we were able to find that a model developed during one season was yielding poor performance for a different season that has completely different temperatures and fleeting reflectors. A user should be able to collect data required for \Name\xspace in 4 hours including topology setup and data collection. The TAO-trees used in \Name\xspace can be trained with CPUs within a few minutes. Since the effort for deploying \Name\xspace is reasonable, it should be easier to develop a model for each season, that can be used for the upcoming years. In future work, we plan to add a more detailed analysis of the retraining gap and the quality of estimation.

If \Name\xspace is deployed in deserts where the temperature is very high during the day and very low during the night with predictable temperature increases and decreases, \Name\xspace can be equipped with two models, one for day time and another for night time. \Name\xspace can then adapt between these models based on the readings from a temperature sensor. If \Name\xspace is deployed in a region where the temperature changes drastically, like the islands of Bora Bora and Maldives, the data collection, training and deployment effort for any ML model will be significant, including \Name. It is hard to design a robust ML-based system for such environments irrespective of the effort made for data collection, training and deployment.

\textbf{Model Adaptation:} Model adaptation is a well-known technique for ML systems to utilize the ML model developed for one location to build the ML model for the other location with reduced data collection and training effort. This is inevitable for training large ML models like LLM's or autonomous driving that require significant time and resources for training a model. However, the effort and resources required for \Name\xspace are minimal. So, \Name\xspace does not require model adaptation. Each location can train \Name's model to accurately reflect the characteristics of the deployed environment.

\textbf{Multi-channel Zigbee is detrimental for \Name.} The drop in E2E throughput performance from 100-300m from the gateway shown in Figure \ref{fig:tput_dist} happens because 3 nodes are contending in a single channel. One could argue that using multi-channel Zigbee will improve the throughput. In general, using multi-channel Zigbee might reduce contention delays and improve throughput for applications that does not focus on throughput and latency. The reason we did not choose multi-channel Zigbee are two-folds: (i) Channel switching needs the transmitting radio to switch to another channel and the receiving radio should also switch to the same channel to receive and process packets. This is could lead to delays. In worst case, it will be similar to the three-way handshake during the radio switching of Gummeson et al. \cite{gummeson2009adaptive} (refer \S\ref{sec:eval}). (ii) \Name\xspace needs fine-grained estimation for the multi-hop Zigbee radios and our deployments uses the $\alpha$=10. This means a node switching channels should send 10 beacon packets to obtain the fine-grained information of the channels quality,  each and every time during a channel switch. The above-mentioned delays combined together will cause a significant delay that will diminish the throughput performance significantly. For mesoscale applications like P2P energy trade in smart-homes where multiple nodes has to bid and respond quickly to a bid, similar to a stock market transaction, the above-mentioned delays will diminish the application performance. So, we decided not to use multi-channel Zigbee. 

\section{Future Work and Conclusion}
This work represents the initial foray into multi-radio architectures for mesoscale IoT applications.  There are many aspects we would like to continue developing.  On the modeling aspect, we are experimenting with TAO-optimized oblique trees for radio selection. While the accuracy could be further improved, the benefits are more related to tree stability when adding additional training data since with axis-aligned trees the tree structure can change dramatically.  This can help gain an understanding of the fundamental properties of the system.  In addition, we would also like to explore different input features for the decision trees.  In particular, we would like to try to feed raw data packet sequences instead of path-quality estimation metrics, and let the model find patterns not easily detected by a human designer on the raw data. We would also like to incorporate the latest LMAC \cite{lmac} for LoRa to enhance the throughput.

To conclude, we presented \Name\xspace a Multi-radio Architecture with Radio Selection using Decision Trees.  The system deploys \Name' nodes in a multi-hop network.  Each of these nodes has two radios, a Zigbee and a LoRa RF transceiver with different throughput/latency features that are optimized for different ranges.  The final goal of \Name\xspace is to select the best radio to be used at any point in the network, using different network paths and link-layer metrics gathered from the radios, to maximize the end-to-end throughput of the data packets being transmitted.  The radio selection is done using novel TAO-optimized decision trees, which are easy to train with limited training data, and easy to deploy in an IoT end device with limited computational power.  In addition, we show that collecting local path metrics as input to our decision trees provides sufficient information to identify the high-throughput radio over the entire path.  \Name\xspace is evaluated on a large-scale complex mesh topology at two different locations. The results show that \Name\xspace can identify the high-throughput radio at the time of transmission. This leads to an average throughput gain of 48.2\% and 49.79\% than the competing schemes, in both locations, while achieving the required latency.

\bibliographystyle{ACM-Reference-Format}
\bibliography{sample-base}

\end{document}